\documentclass[aps,prl,a4paper,preprintnumbers,superscriptaddress,showpacs,tightenlines,amsmath,amssymb]{revtex4}
\usepackage{graphicx,color} % Include figure files
\usepackage{dcolumn}  % Align table columns on decimal point
\usepackage{pstricks}% required
\usepackage{ulem}

\graphicspath{{ps}}

\begin{document}

%\preprint{\vbox{
%                 \hbox{PRD draft}
%                 \hbox{Version 2.0}
%}}
%\vspace*{-3\baselineskip}
%\resizebox{!}{3cm}{\includegraphics{belle.eps}}

\title{ \quad\\[1.0cm] Measurement of $CP$ violating asymmetries 
in $B^0\rightarrow K^+K^- K^0_S$ decays with a time-dependent Dalitz approach}

%%%% >>>>> insert the authorlist here. BEFORE the abstract !!!!! <<<<<
%%%% >>>>> from the authorship confirmation web page
%%% Name the file author.tex and use \input{author} to insert into your latex file.
%%% Paper:    B0 -> K K KS TCPV
%%% Journal:  Physical Review D
%%% Contacts: K. Sumisawa (kazutaka.sumisawa@kek.jp)
%%%           Y. Nakahama (nakahama@hep.phys.s.u-tokyo.ac.jp)
%%% Non-responding authors or those who said NO are commented out.
%%% ====================================================================
%%% Click the RELOAD button on your web browser to see the updated file.
%%% ====================================================================
%%% Use \input{author} to insert this material into your latex file.
%%%%% Force institutions to appear in alphabetical order when typeset.
\affiliation{Budker Institute of Nuclear Physics, Novosibirsk}
\affiliation{Faculty of Mathematics and Physics, Charles University, Prague}
\affiliation{Chiba University, Chiba}
\affiliation{University of Cincinnati, Cincinnati, Ohio 45221}
%%%\affiliation{Department of Physics, Fu Jen Catholic University, Taipei}
%%%\affiliation{Justus-Liebig-Universit\"at Gie\ss{}en, Gie\ss{}en}
%%%\affiliation{The Graduate University for Advanced Studies, Hayama}
%%%\affiliation{Gyeongsang National University, Chinju}
\affiliation{Hanyang University, Seoul}
\affiliation{University of Hawaii, Honolulu, Hawaii 96822}
\affiliation{High Energy Accelerator Research Organization (KEK), Tsukuba}
%%%\affiliation{Hiroshima Institute of Technology, Hiroshima}
%%%\affiliation{University of Illinois at Urbana-Champaign, Urbana, Illinois 61801}
%%%\affiliation{Indian Institute of Technology Guwahati, Guwahati}
\affiliation{Institute of High Energy Physics, Chinese Academy of Sciences, Beijing}
\affiliation{Institute of High Energy Physics, Vienna}
\affiliation{Institute of High Energy Physics, Protvino}
%%%\affiliation{Institute of Mathematical Sciences, Chennai}
%%%\affiliation{INFN - Sezione di Torino, Torino}
\affiliation{Institute for Theoretical and Experimental Physics, Moscow}
\affiliation{J. Stefan Institute, Ljubljana}
\affiliation{Kanagawa University, Yokohama}
\affiliation{Institut f\"ur Experimentelle Kernphysik, Karlsruher Institut f\"ur Technologie, Karlsruhe}
\affiliation{Korea Institute of Science and Technology Information, Daejeon}
\affiliation{Korea University, Seoul}
%%%\affiliation{Kyoto University, Kyoto}
\affiliation{Kyungpook National University, Taegu}
\affiliation{\'Ecole Polytechnique F\'ed\'erale de Lausanne (EPFL), Lausanne}
\affiliation{Faculty of Mathematics and Physics, University of Ljubljana, Ljubljana}
\affiliation{University of Maribor, Maribor}
\affiliation{Max-Planck-Institut f\"ur Physik, M\"unchen}
\affiliation{University of Melbourne, School of Physics, Victoria 3010}
\affiliation{Nagoya University, Nagoya}
%%%\affiliation{Nara University of Education, Nara}
\affiliation{Nara Women's University, Nara}
\affiliation{National Central University, Chung-li}
\affiliation{National United University, Miao Li}
\affiliation{Department of Physics, National Taiwan University, Taipei}
\affiliation{H. Niewodniczanski Institute of Nuclear Physics, Krakow}
\affiliation{Nippon Dental University, Niigata}
\affiliation{Niigata University, Niigata}
\affiliation{University of Nova Gorica, Nova Gorica}
\affiliation{Novosibirsk State University, Novosibirsk}
\affiliation{Osaka City University, Osaka}
%%%\affiliation{Osaka University, Osaka}
\affiliation{Panjab University, Chandigarh}
%%%\affiliation{Peking University, Beijing}
%%%\affiliation{Princeton University, Princeton, New Jersey 08544}
%%%\affiliation{RIKEN BNL Research Center, Upton, New York 11973}
%%%\affiliation{Saga University, Saga}
\affiliation{University of Science and Technology of China, Hefei}
\affiliation{Seoul National University, Seoul}
%%%\affiliation{Shinshu University, Nagano}
\affiliation{Sungkyunkwan University, Suwon}
\affiliation{School of Physics, University of Sydney, NSW 2006}
\affiliation{Tata Institute of Fundamental Research, Mumbai}
\affiliation{Excellence Cluster Universe, Technische Universit\"at M\"unchen, Garching}
\affiliation{Toho University, Funabashi}
\affiliation{Tohoku Gakuin University, Tagajo}
\affiliation{Tohoku University, Sendai}
\affiliation{Department of Physics, University of Tokyo, Tokyo}
\affiliation{Tokyo Metropolitan University, Tokyo}
\affiliation{Tokyo University of Agriculture and Technology, Tokyo}
%%%\affiliation{Toyama National College of Maritime Technology, Toyama}
\affiliation{IPNAS, Virginia Polytechnic Institute and State University, Blacksburg, Virginia 24061}
\affiliation{Wayne State University, Detroit, Michigan 48202}
\affiliation{Yonsei University, Seoul}
% \author{I.~Adachi}\affiliation{High Energy Accelerator Research Organization (KEK), Tsukuba} % KEK
  \author{Y.~Nakahama}\affiliation{Department of Physics, University of Tokyo, Tokyo} % Tokyo
  \author{K.~Sumisawa}\affiliation{High Energy Accelerator Research Organization (KEK), Tsukuba} % KEK
  \author{H.~Aihara}\affiliation{Department of Physics, University of Tokyo, Tokyo} % Tokyo
  \author{K.~Arinstein}\affiliation{Budker Institute of Nuclear Physics, Novosibirsk}\affiliation{Novosibirsk State University, Novosibirsk} % BINP
% \author{T.~Aso}\affiliation{Toyama National College of Maritime Technology, Toyama} % Toyama
% \author{V.~Aulchenko}\affiliation{Budker Institute of Nuclear Physics, Novosibirsk}\affiliation{Novosibirsk State University, Novosibirsk} % BINP
  \author{T.~Aushev}\affiliation{\'Ecole Polytechnique F\'ed\'erale de Lausanne (EPFL), Lausanne}\affiliation{Institute for Theoretical and Experimental Physics, Moscow} % ITEP
  \author{T.~Aziz}\affiliation{Tata Institute of Fundamental Research, Mumbai} % Tata
  \author{A.~M.~Bakich}\affiliation{School of Physics, University of Sydney, NSW 2006} % Sydney
  \author{V.~Balagura}\affiliation{Institute for Theoretical and Experimental Physics, Moscow} % ITEP
% \author{Y.~Ban}\affiliation{Peking University, Beijing} % Peking
% \author{E.~Barberio}\affiliation{University of Melbourne, School of Physics, Victoria 3010} % Melbourne
% \author{A.~Bay}\affiliation{\'Ecole Polytechnique F\'ed\'erale de Lausanne (EPFL), Lausanne} % Lausanne
% \author{I.~Bedny}\affiliation{Budker Institute of Nuclear Physics, Novosibirsk}\affiliation{Novosibirsk State University, Novosibirsk} % BINP
  \author{K.~Belous}\affiliation{Institute of High Energy Physics, Protvino} % Protvino
  \author{V.~Bhardwaj}\affiliation{Panjab University, Chandigarh} % Panjab
% \author{B.~Bhuyan}\affiliation{Indian Institute of Technology Guwahati, Guwahati} % IITG
% ...
% Karlsruhe
% \author{N.~Zwahlen}\affiliation{\'Ecole Polytechnique F\'ed\'erale de Lausanne (EPFL), Lausanne} % Lausanne
% \author{O.~Zyukova}\affiliation{Budker Institute of Nuclear Physics, Novosibirsk}\affiliation{Novosibirsk State University, Novosibirsk} % BINP
\collaboration{The Belle Collaboration}

\begin{abstract}
We report a measurement of $CP$ violating asymmetries in
$B^0(\overline{B}^0) \to K^+ K^- K^0_S$ decays with a time-dependent Dalitz approach.
This analysis is based on a data sample of $657\times 10^6$ $B\overline{B}$ pairs 
accumulated at the $\Upsilon(4S)$ resonance with the Belle detector at 
the KEKB asymmetric-energy $e^+e^-$ collider.
As the result of an unbinned maximum likelihood fit to the selected candidates,
the mixing-induced and direct $CP$ violation parameters, $\phi^{\rm eff}_1$
and ${\cal A}_{CP}$ are obtained for $B^0 \to \phi(1020) K^0_S$, $B^0 \to f_0(980) K^0_S$
and other $B^0 \to K^+ K^- K^0_S$ decays. We find four solutions that describe the data.
There are
\begin{eqnarray*}
\phi_1^{\rm eff}(B^0\to \phi(1020) K^0_S) & = & (32.2 \pm 9.0 \pm 2.6 \pm 1.4)^{\circ};\\
\phi_1^{\rm eff}(B^0\to \phi(1020) K^0_S) & = & (26.2 \pm 8.8 \pm 2.7 \pm 1.2)^{\circ};\\
\phi_1^{\rm eff}(B^0\to \phi(1020) K^0_S) & = & (27.3 \pm 8.6 \pm 2.8 \pm 1.3)^{\circ}\; {\rm and}\\
\phi_1^{\rm eff}(B^0\to \phi(1020) K^0_S) & = & (24.3 \pm 8.0 \pm 2.9 \pm 5.2)^{\circ}.
\end{eqnarray*}
The values for the $CP$ violating phase in $B^0\to \phi(1020) K^0_S$ are similar but
other properties of the Dalitz plot are quite different for the four solutions.
These four solutions have consistent $\phi^{\rm eff}_1$ values for all three
$B$ meson decay channels and none of them deviates significantly from
the values measured in $B \to (c\bar{c}) K^0$ decays with the currently available
statistics.
In addition, we find no significant direct $CP$ violation.
\end{abstract}

\pacs{12.15.Hh, 13.25.Hw}

\maketitle

%%%% >>>> keep the final version single-spaced
\tighten

{\renewcommand{\thefootnote}{\fnsymbol{footnote}}}
\setcounter{footnote}{0}

%${\bf Introduction}$
$CP$ violation in the quark sector is described in the standard model (SM) by
the Kobayashi-Maskawa (KM) theory~\cite{KM}.
In this theory, the existence of a single irreducible phase gives rise to
$CP$ violating asymmetries in the time-dependent rates 
of $B^0$ and $\overline{B}^0$ decays 
into a common $CP$ eigenstate~\cite{carter}.
Specifically, for neutral $B$ meson decays dominated by
$b \rightarrow c\bar{c}s$ transitions such as 
$B^0(\overline{B}^0) \rightarrow J/\psi K^0$,
we can measure $CP$ violating quantities that determine
the $\phi_1$~\cite{phi1_beta} angle of the Unitarity Triangle.
%Theoretically very clean measurements have been performed by
%Belle~\cite{jpsiks_Belle} and BaBar~\cite{jpsiks_BABAR} collaborations,
%and provide a precise reference value for $\phi_1$.
The measurements have been performed by
Belle~\cite{jpsiks_Belle} and BaBar~\cite{jpsiks_BABAR} collaborations,
and provide a precise reference value for $\phi_1$ because of the very small theoretical uncertainty.

Recently, measurements of time-dependent $CP$ violation of
$b \rightarrow s$ penguin-mediated $B$ decays
have become interesting because these decay modes
proceed via loop diagrams and are, therefore, expected
to be sensitive probes of the physics beyond the SM.
In these decay modes, searches for new physics
effects are carried out by investigating deviations of $CP$
violating parameters
from those determined by $b \rightarrow c\bar{c}s$ processes~\cite{b2sTheory}.

Among these $B^0$ decays,  $B^0 \rightarrow K^+K^-K^0_S$ is
one of the most promising modes because of its very small Cabibbo-suppressed 
tree diagram contribution.
Previous Belle measurements of the $CP$ violating asymmetries 
have been performed separately in the $K^+ K^-$ mass region around 
the $\phi(1020)$ mass~\cite{ref:belle_phiks_cp} and at
higher $K^+ K^-$ masses~\cite{ref:belle_kskk_cp}, while
neglecting interference between intermediate states.
It is, however, expected that the sensitivity to $CP$ violating parameters would
improve in a measurement using the time-dependent Dalitz plot distribution
because of the correct treatment of interferences between various 
resonant and nonresonant $B^0 \rightarrow K^+K^-K^0_S$ processes.

In the decay chain
$\Upsilon$(4S)$\to B^0\overline{B}^0\to$ $(K^+K^-K^0_S) f_{\rm tag}$, 
where one of the $B$ mesons decays at time $t_{\rm rec}$ to 
the final state $K^+K^-K^0_S$ and 
the other decays at time $t_{\rm tag}$ to a final state $f_{\rm tag}$ 
that distinguishes between $B^0$ and $\overline{B}^0$,
the decay rate has a time dependence given by%~\cite{cpviolation}
\begin{eqnarray}
\label{signal_cp}
%KM |\mathcal{A}(\Delta{t},q)|^2 
|A_{\rm sig}(\Delta{t},q)|^2 
& = & \frac{ e^{-|\Delta{t}|/{\tau_{B^0}}} }{4\tau_{B^0}}
 \Bigl[ (|A|^2 + |\bar{A}|^2) \\ \nonumber
& - & q(|A|^2 - |\bar{A}|^2)\cos(\Delta m_d \Delta{t}) \\ \nonumber
& + & 2q{\mathcal Im}(\bar{A}A^*)\sin(\Delta m_d \Delta{t}) 
\Bigr],
\end{eqnarray}
where $\tau_{B^0}$ is 
the neutral $B$ meson lifetime,
$\Delta m_d$ is the mass difference between 
the two neutral $B$ mass eigenstates, 
$\Delta t = t_{\rm rec}-t_{\rm tag}$, 
$A(\bar{A})$ is the total amplitude of 
$B^0(\overline{B}^0) \rightarrow K^+K^- K^0_S$, 
and the $b$-flavor charge $q = +1~(-1)$ when the tagged $B$ meson is 
a $B^0(\overline{B}^0)$.
The Dalitz plot variables $s_+$, $s_-$, and $s_0$ are defined as 
$s_{\pm} \equiv (p_{\pm} + p_0)^2$, and $s_{0} \equiv (p_{+} + p_{-})^{2}$, 
where $p_+$, $p_-$, and $p_0$ are the four-momenta of the $K^+$, $K^-$,
and $K^0_S$, respectively.
These variables satisfy $s_+ + s_- + s_0 = m^2_{B^0} +2m^2_{K^+} + m^2_{K^0_S}$ by energy-momentum conservation.
In the isobar approximation~\cite{ref:isobar}, the total amplitude for
$B^{0}(\overline{B}^0) \rightarrow K^{+} K^{-} K^{0}_{S}$ is given by
 the sum of the decay channels with that final state,
\begin{equation}
  A(s_{+}, s_{-}) =\sum_{i} a^{\prime}_{i}F_{i}(s_{+}, s_{-}), \;\;\;\;
 \bar A(s_{-}, s_{+}) =\sum_{i} \bar a^{\prime}_{i} \bar F_{i}(s_{-}, s_{+}),
\label{eq:kskkDecayAmplitude}
\end{equation}
where $a_{i}^{\prime} \equiv a_{i}e^{ib_{i}}$ is a complex coefficient
describing the relative magnitude and phase for the $i$-th decay
channel, including the weak phase dependence.
The Dalitz-dependent amplitudes, $F_{i}(s_{+}, s_{-})$, contain only
strong dynamics and, thus, $F_{i}(s_{+}, s_{-}) = \bar F_{i}(s_{-},s_{+})$.
The amplitudes of the contributions considered in the $B^0 \rightarrow K^+K^-K^0_S$ decay 
are summarized in Table~\ref{tab_sigmod}.
%We use the same formalism as the Belle $B^0 \rightarrow \pi^+\pi^-K^0_S$ 
%time-dependent Dalitz analysis~\cite{belle_pipiks_dalitz} 
%with the pions replaced by charged kaons in
%the $K^+K^-K^0_S$ final state.
We use the same formalism as references~\cite{belle_pipiks_dalitz, babar_kkks_dalitz}.
We utilize Flatt\'{e}~\cite{ref:Flatte} and 
relativistic Breit Wigner (RBW)~\cite{ref:pdg} lineshapes
to describe the resonances.
%taking Blatt-Weiskopf barrier factors~\cite{ref:BWBF} to be unity.

%%%%%% Table 1 %%%
%\begin{table}[htb]
\begin{table}
\caption{Summary of the contributions in the signal model. 
%KM $L$ is e angular momentum.
Here $L$ is the orbital angular momentum.
All fixed parameters are taken from Ref.~\cite{ref:pdg} except for 
those of the $f_0(980)$~\cite{ref:flattetwo} and $f_{\rm X}$~\cite{ref:belle_kkk}.
}
\label{tab_sigmod}
\begin{tabular}{cccc}
\hline \hline
Resonances & Fixed parameters (MeV) & Resonance shape & $L$\\ 
\hline
$f_0(980)$   & $M$ = 965$\pm$10 &  Flatt\'{e} & 0\\
         & $g_{\pi}$=165$\pm$18 &    \\
         & $g_{K}$=(4.21$\pm$0.33)$g_{\pi}$ &  &  \\
$\phi(1020)$ & $M$ = 1019.455$\pm$0.020 &  RBW & 1\\
         & $\Gamma$ = 4.26$\pm$0.04 &   \\
$f_{\rm X}$    & $M$ = 1524$\pm$14  &  RBW  & 0\\
         & $\Gamma$ = 136$\pm$23 &   \\
$\chi_{c0}$ & $M$ = 3414.75$\pm$0.31 & RBW & 0 \\
         & $\Gamma$ = 10.4$\pm$0.7 &   \\
$(K^+K^-)_{\rm NR}$    & no fixed parameters & $e^{-\alpha s_0}$ &  \\
$(K^+K^0_S)_{\rm NR}$  & no fixed parameters  & $e^{-\alpha s_+}$ &  \\
$(K^-K^0_S)_{\rm NR}$  & no fixed parameters  & $e^{-\alpha s_-}$ &  \\
\hline \hline
\end{tabular}
\end{table}
%%%%%%%%%%

In the Dalitz-dependent amplitudes, $A$ (Eq.~\ref{eq:kskkDecayAmplitude}), we choose a convention
in which the $B^0 \overline{B}^0$ mixing
% phase of $q/p$
phase ($q/p$) is absorbed into the $\overline{B}^0$ decay
amplitude, $\bar a_{i}^{\prime}$.
These complex coefficients, $a^{\prime}_{i}$ and $\bar a^{\prime}_{i}$,
can be redefined as
\begin{equation}
  a^{\prime}_{i} \equiv a_{i}(1 + c_{i})e^{i(b_{i} + d_{i})}, \;\;\;\;
  \bar a^{\prime}_{i} \equiv a_{i}(1 - c_{i})e^{i(b_{i} - d_{i})},
\label{eq:resonance_ampl}
\end{equation}
in which case
a resonance, $i$, has a direct $CP$ violating asymmetry given by
\begin{equation}
  {\cal A}_{CP}(i) \equiv \frac{|\bar a^{\prime}_{i}|^{2}-|a^{\prime}_{i}|^{2}}{|\bar a^{\prime}_{i}|^{2}+|a^{\prime}_{i}|^{2}} =  \frac{-2c_{i}}{1+c_{i}^{2}},
\label{eq:acp_dtcpv}
\end{equation}
where the $c_i$'s are restricted by definition to lie between $-1$ and $1$.

For cases where the contribution $i$ is a $CP$ eigenstate, 
the mixing-induced $CP$ violating parameter, $\phi^{\rm eff}_1(i)$,
equals the fitted parameter $d_{i}$,
\begin{equation}
  \phi^{\rm eff}_1(i)
 \equiv \frac{\arg(a^{\prime}_{i} \bar a^{\prime *}_{i})}{2}
 = d_{i},
\label{eq:phione_dtcpv}
\end{equation}
and is related to the mixing-induced  $CP$ violating asymmetry as
\begin{equation}
   -\eta_{i}{\cal S}(i)
 \equiv \frac{-2{\rm Im}(\bar a^{\prime}_{i}a^{\prime *}_{i})}{|a^{\prime}_{i}|^{2} + |\bar a^{\prime}_{i}|^{2}}
 = \frac{1-c^{2}_{i}}{1+c^{2}_{i}}\sin 2 \phi^{\rm eff}_1(i),
\label{eq:dtcpv_sin2phione}
\end{equation}
where $\eta_{i}$ is the $CP$ eigenvalue of the final state. 
%KM Note that ${\cal A}(i)$ and ${\cal S}(i)$ are restricted 
Note that ${\cal A}_{CP}(i)$ and ${\cal S}(i)$ are restricted 
by these definitions to lie in the physical region.

%${\bf KEKB and Belle Detector}$
This time-dependent Dalitz measurement of $CP$ violating parameters in 
$B^0 \rightarrow K^+K^-K^0_S$ decays is based on a large data 
sample that contains $657\times 10^6$ $B\overline{B}$ pairs, 
collected  with the Belle detector at the KEKB asymmetric-energy
$e^+e^-$ (3.5 on 8~GeV) collider~\cite{KEKB}
operating at the $\Upsilon(4S)$ resonance.
The $\Upsilon(4S)$ is produced with a Lorentz boost factor
of $\beta\gamma = 0.425$ along the $z$-axis, 
which is antiparallel to the positron beam direction.
Since the $B\overline{B}$ pairs are produced nearly at rest
in the $\Upsilon(4S)$ center-of-mass system (cms),
$\Delta t$ is determined from $\Delta z$, the distance 
between the two $B$ meson decay vertices along 
the $z$-direction: $\Delta t \cong \Delta z/c\beta\gamma$, 
where $c$ is the speed of light.

The Belle detector is a large-solid-angle magnetic
spectrometer that
consists of a silicon vertex detector (SVD),
a 50-layer central drift chamber (CDC), an array of
aerogel threshold Cherenkov counters (ACC), 
a barrel-like arrangement of time-of-flight
scintillation counters (TOF), and an electromagnetic calorimeter
comprised of CsI(Tl) crystals (ECL) located inside 
a superconducting solenoid coil that provides a 1.5~T
magnetic field.  An iron flux-return located outside
the coil is instrumented to detect $K^0_L$ mesons and to identify
muons (KLM).  
The detector is described in detail elsewhere~\cite{Belle}.
Two inner detector configurations were used. 
A 2.0 cm radius beam pipe
and a 3-layer silicon vertex detector were used for the first sample
of $152\times 10^6$ $B\overline{B}$ pairs, 
while a 1.5 cm radius beampipe, a 4-layer
silicon detector and a small-cell inner drift chamber were used to record  
the remaining $505\times 10^6$ $B\overline{B}$ pairs~\cite{svd2}.

%${\bf event selection}$
We reconstruct $B$ candidates from an oppositely-charged kaon pair and a  $K^0_S$ candidate.
The charged kaons are selected from the charged tracks
having their impact parameters consistent with coming from the interaction 
point (IP). 
%Particle identification (PID) is determined with the likelihood ratio, ${\cal R}_{i/j} = P_i/(P_i+P_j)$,
%where $P_i(P_j)$ is the likelihood that the particle is type $i$($j$).
To suppress background from particle misidentification, charged tracks
that are positively identified as pions, protons, or electrons are excluded.
The particle species are identified by using particle information (PID) from the CDC, ACC, TOF, and ECL systems.
We reconstruct $K^0_S$ candidates from pairs of oppositely charged tracks having invariant mass within 
12 MeV/$c^2$ of the $K^0_S$ mass.
The direction of the $K^0_S$ momentum is required to be consistent with 
the direction of vertex displacement with respect to IP~\cite{ref:belle_b2sqq_2005}.
%The distance of closest approach of the candidate charged tracks to the IP in the plane perpendicular
%to $z$ axis is required to be large than 0.02 cm for high-momentum ($> 1.5$ GeV/$c$) $K^0_S$ candidates
%and 0.03 cm for those with momentum less than 1.5 GeV/$c$.
%The $\pi^+ \pi^-$ vertex is required be displaced from the IP by a minimum transverse distance of 0.22 cm
%for high-momentum candidates and 0.08 cm for the remaining candidates.
%The mismatch in the $z$ direction at the $K^0_S$ vertex point for the $\pi^+ \pi^-$ tracks must be less than 2.4 cm
%for high-momentum candidates and 1.8 cm for the remaining candidates.
%The direction of the pion pair momentum must agree also with the direction of the vertex point from the IP to within 0.03 rad
%for high-momentum candidates and to within 0.03 rad for the remaining candidates.

We combine the $K^+K^-$ pair and $K^0_S$ to form a neutral $B$ meson. 
Signal candidates are identified by two kinematic variables defined
in the cms: the beam-energy constrained mass 
$M_{\rm bc}\equiv \sqrt{E^{2}_{\rm beam}-(\overrightarrow{p}_B)^2}$ 
and the energy difference $\Delta E \equiv E_B - E_{\rm beam}$, 
where $E_{\rm beam}=\sqrt{s}/2$ is the cms beam energy, 
and $\overrightarrow{p}_B$ and $E_B$ are the cms three momentum
and energy of the reconstructed $B$ meson candidate, respectively.
We use candidates in a signal region defined as a $3\sigma$ ellipse around the
$M_{\rm bc}$ and $\Delta E$ mean values:
$\frac{(M_{\rm bc} -\ensuremath{M_{B^0}})^2}{(8\;{\rm MeV}/c^2)^2}
+\frac{(\Delta E)^2}{(45\; {\rm MeV})^2}<1$,
where $M_{B^0}$ is the nominal neutral $B$ meson mass~\cite{ref:pdg}.
A larger region in $M_{\rm bc}$ and $\Delta E$,
 $5.20$ GeV/$c^2 < M_{\rm bc}$ and $-0.30$ GeV $< \Delta E <$ $0.50$ GeV,
is used to determine the signal and background fractions.
The sideband regions used for the continuum background study 
are defined as
$5.20$ GeV/$c^2 < M_{\rm bc} < 5.26$ GeV/$c^2$ and $-0.10$ GeV $<\Delta E<0.50$ GeV for the $\Delta t$ distribution, 
$5.24$ GeV/$c^2 < M_{\rm bc} < 5.30$ GeV/$c^2$ and $-0.10$ GeV $<\Delta E<0.10$ GeV excluding the rectangular region of
$5.268$ GeV/$c^2 < M_{\rm bc} < 5.30$ GeV/$c^2$ and $-0.05$ GeV $<\Delta E<0.05$ GeV 
for the Dalitz distribution.

The dominant source of background is continuum
$e^+e^- \rightarrow q\bar{q}$ $(q=u,d,s,$ and $c)$ production.
To reduce it, we require that $|\cos \theta_{\rm th}| < 0.8$,
where $\theta_{\rm th}$ is the angle between the thrust axis of the $B$ candidate and that of the rest of the event.
%and the $B^0 \rightarrow K^+K^-K^0_S$ decay candidate momentum.
This requirement retains 83\% of the signal while 79\% of the continuum events are removed.
%The dominant $B \overline{B}$ background is found to originate from
The $B \overline{B}$ background is found to be mostly originating from
$b\to c$ $B$-decays, which peaks in the signal region with an estimated yield of $60$ events:
$B^0 \to D^- [K^0_S K^{-}] K^{+}$~\cite{charge_conj} and $B^0 \to J/\psi K^0_S$ decays.
There are also potential backgrounds from $B^0 \to D_s^-[K^0_S K^-] K^+$
and $B^0 \to {\overline{D}^0}[K^+K^-] K^0_S$.
Backgrounds due to $K$--$\pi$ misidentification are also found.
All these peaking background decays are suppressed to a negligible level
by applying $\sim1.5 \sigma\ (J/\psi)$ and $\sim2.5 \sigma$ (other modes) vetoes on the invariant masses;
these vetoes are summarized in Table~\ref{table:charm_veto}.
For backgrounds that arise from misidentified particles, the invariant masses are
recalculated by assuming an alternate mass hypothesis for the charged kaon.
The remaining contribution is included in the nominal fit as the $B \overline{B}$ background component.
Yields for signal, continuum and $B \overline{B}$ backgrounds
as well as PDFs for those are described in more detail later.

%\begin{table}[htbp]
\begin{table}
\begin{center}
\caption{Summary of the charm vetoes applied to $B^0 \rightarrow K^+K^-K^0_S$ candidates. The subscript in the vetoed region indicates that an alternate mass hypothesis has been applied to the kaon candidates used to calculate the invariant mass term.}
\begin{tabular}{cc} \hline\hline 
Vetoed mode & Vetoed region \\ \hline
$B^0 \to D^- [K^0_S K^{-}]K^+$		    & $|M(K^0_S K^{-})-M_{D^{-}}|<15\;{\rm MeV}/c^2$               \\
$B^0 \to J/\psi [K^+K^-] K^0_S$ 	    & $|M(K^{+}K^{-})-M_{J/\psi}|<15\; {\rm MeV}/c^2$              \\
$B^0 \to D_s^-[K^0_S K^-] K^+$		    & $|M(K^0_S K^{-})-M_{D_{s}^-}|<15\;{\rm MeV}/c^2$             \\
$B^0 \to {\overline{D}^0}[K^+K^-] K^0_S$    & $|M(K^{+}K^{-})-M_{\overline{D}^0}|<15\;{\rm MeV}/c^2$       \\
$B^0 \to D^- [K^0_S \pi^{-}]K^+$	    & $|M(K^0_S K^{-})_{\pi}-M_{D^{-}}|<15\;{\rm MeV}/c^2$         \\
$B^0 \to {\overline{D}^0}[ K^+\pi^-] K^0_S$ & $|M(K^{+}K^{-})_{\pi}-M_{\overline{D}^0}|<15\;{\rm MeV}/c^2$ \\
\hline\hline
\end{tabular}
\label{table:charm_veto}
\end{center}
\end{table}

We identify the flavor of the accompanying $B$ meson 
from inclusive properties of particles
that are not associated with the reconstructed $B^0 \rightarrow K^+K^-K^0_S$ candidate.
The algorithm for flavor tagging is described 
in detail elsewhere~\cite{tagging}.
To represent the tagging information, we use two parameters, $q$ defined in Eq.(\ref{signal_cp}) and $r$.
The parameter $r$ is an event-by-event Monte Carlo (MC) determined 
flavor-tagging quality factor that ranges from $r=0$ for no flavor 
discrimination to $r=1$ for unambiguous flavor assignment. 
It is used only for sorting data into seven intervals. 
The wrong tag fractions for the seven $r$ intervals, $w_l$ $(l=0,6)$, 
and the difference in $w$ between $B^0$ and $\overline{B}{}^0$ decays, 
$\Delta w_l$, are determined from data~\cite{tagging}.
The vertex position for the $K^+K^-K^0_S$ decay is reconstructed using
the charged kaon pair and the transverse components of IP.
The vertex position of $f_{\rm tag}$ is obtained using tracks that
are not assigned to the $K^+K^-K^0_S$ candidate and IP.

We find that 1.5\;\% of the selected events have more than one
$B^0 \rightarrow K^+K^-K^0_S$ candidate.
In these events, we choose the $B^0$ candidate
that is formed from the most kaon-like charged kaon candidate
%by requiring the highest kaon ID likelihood ratio (${\cal R}_{K/\pi}$)
and the $K^0_S$ candidate closest to the nominal $K^0_S$ mass.
%and the smallest (${\Delta M_{K^0_S}}$)$^2$ value.

\begin{figure}
\begin{center}
\includegraphics[width=0.3\textwidth]{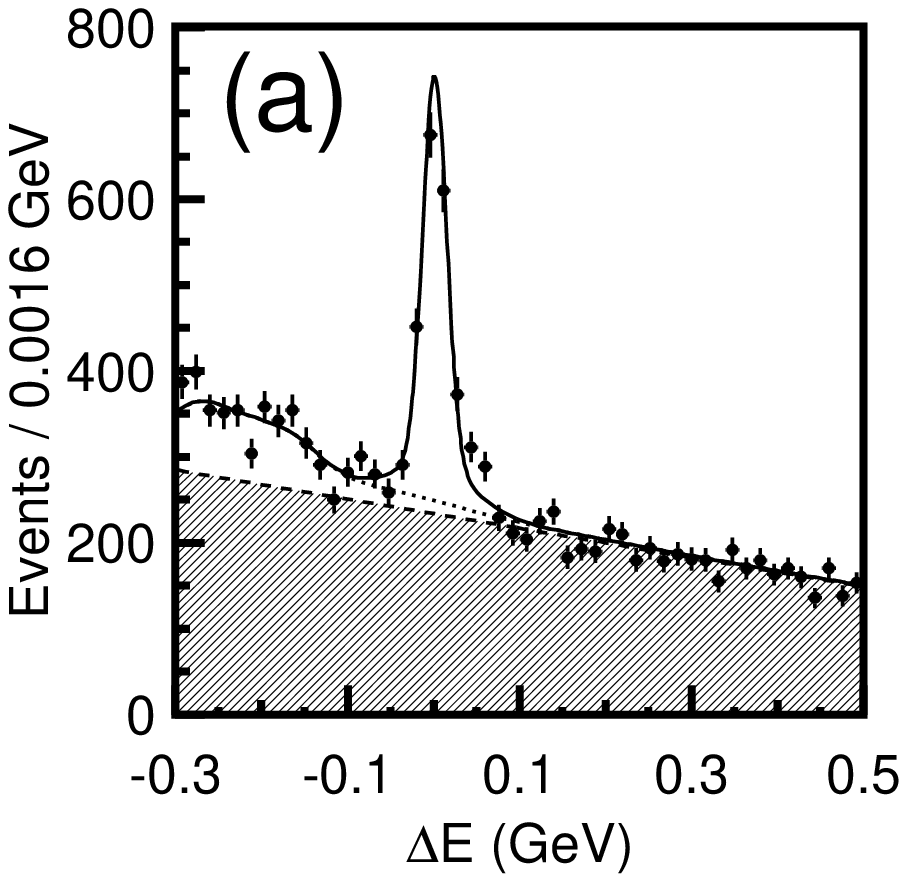} 
\includegraphics[width=0.3\textwidth]{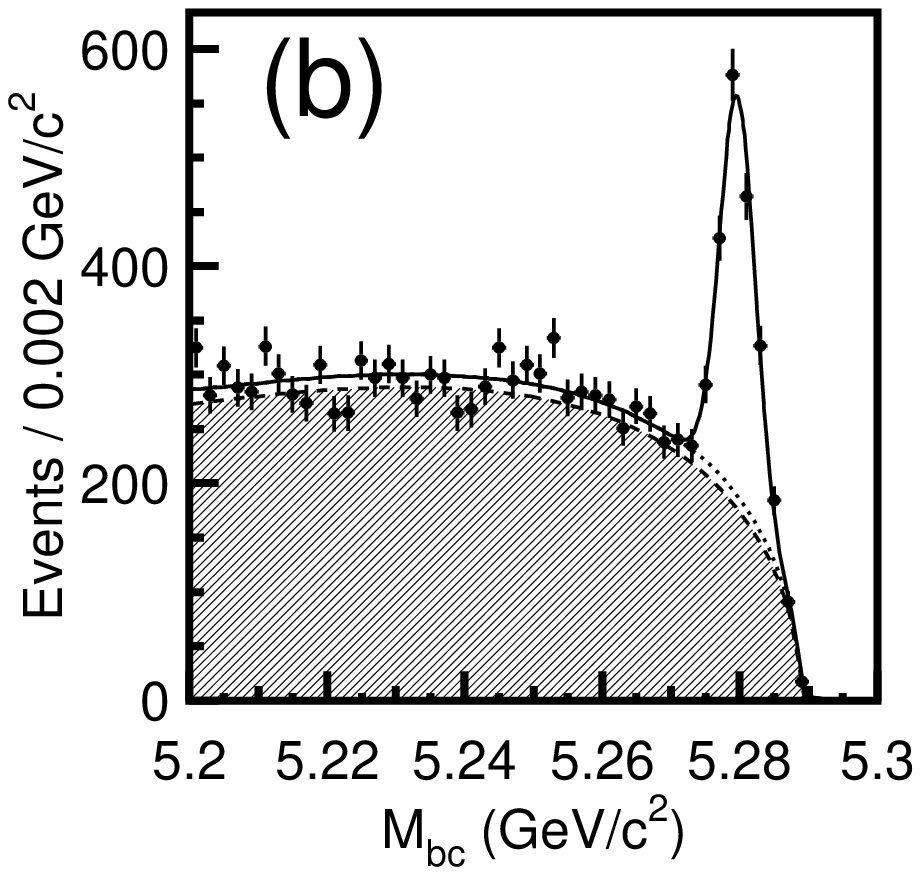}
\includegraphics[width=0.3\textwidth]{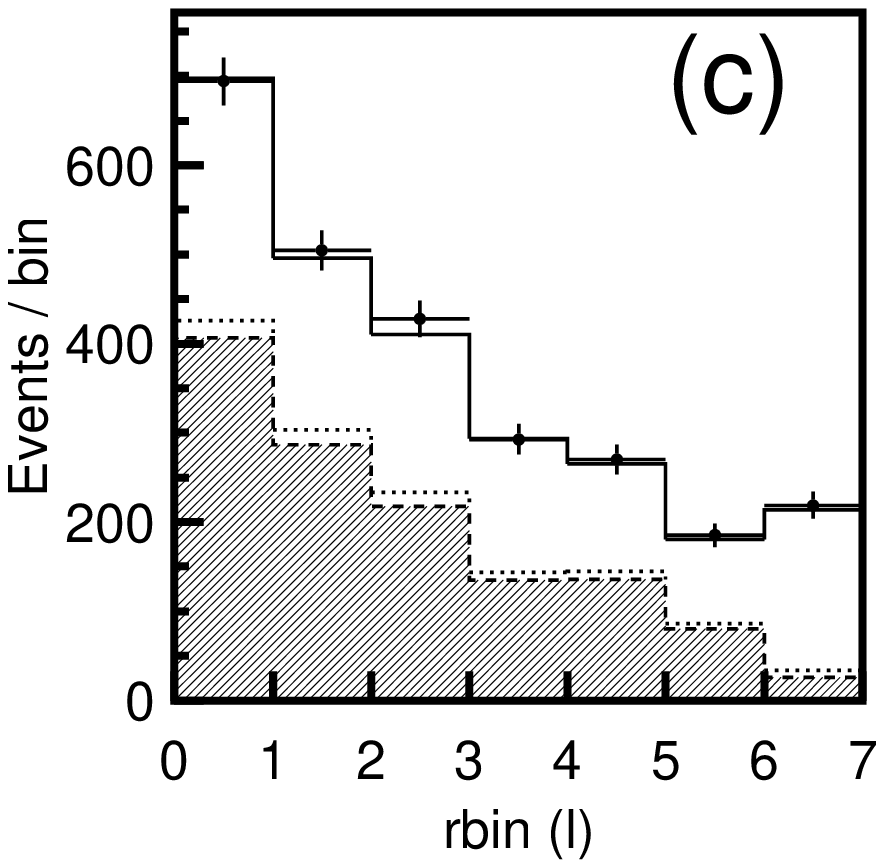}  
\caption{Signal enhanced total projections of (a) $\Delta E$ with $5.272\;$GeV/$c^2<M_{\rm bc}<5.288\;$GeV/$c^2$, (b) $M_{\rm bc}$ with $|\Delta E|<0.045\;{\rm GeV}$ and (c) $l$ in the ($\Delta E$,$M_{\rm bc}$) signal region for the $B^0 \to K^+K^-K^0_S$ candidate events. The solid curves show the fit projections, the hatched areas show the continuum background component and the dotted curves show the total background contribution. The points with error bars are the data.} 
\label{fig:fig_yield}
\end{center}
\end{figure}

%${\bf signal yield}$
After all the selections are applied, we obtain 98982 candidates in the
$M_{\rm bc}$-$\Delta E$ fit region, of which 2333 are in the signal region.
We extract the signal yield using a three-dimensional extended unbinned
maximum likelihood fit to the distributions of
$\Delta E$, $M_{\rm bc}$ and the flavor-tag quality ($r$) interval,
$l$, for the selected $B^0 \rightarrow K^+K^-K^0_S$ events.
For the probability density function (PDF) of the signal component,
we use a sum of two Gaussians (a single Gaussian) for the $\Delta E$
($M_{\rm bc}$) shape.
All parameters of the PDFs are free in the fit,
except the ratio of the area of the broader Gaussian component to that of the core Gaussian,
 and the width of the broader Gaussian in $\Delta E$.
These additional parameters are fixed from the results of a fit to a
$B^0 \to D^-[K^0_S \pi^-]\pi^+$ data control sample.
For the continuum background component,
the $\Delta E$~($M_{\rm bc}$) shape is modeled by a first-order
polynomial (an ARGUS~\cite{argus}) function,
with shape parameters floated in the fit.
The $B \overline{B}$ background component is parameterized by
two-dimensional binned histograms from MC.
In the fit, the total signal,
continuum and $B \overline{B}$ background yields are also free parameters.
The fit yields $1176\pm 51$ signal events
in the signal region.
The projections of the $\Delta E$, $M_{\rm bc}$ and
$l$ distributions for the candidate events are shown in
Fig.~\ref{fig:fig_yield}.
The average signal, continuum and $B\overline{B}$ fractions in 
the signal ellipse are calculated to be 
$\sim$50\;\%, $\sim$49\;\% and $\sim$1\;\%, respectively.
The event-by-event signal probabilities as a function of $\Delta E$, $M_{\rm bc}$ and $l$ obtained with this
fit are used in the unbinned maximum likelihood fit
with a time-dependent Dalitz approach that is used to extract $CP$ violation parameters.

%\begin{figure}[htb]

\begin{figure}
\begin{center}                                                             
\includegraphics[height=14cm,angle=270]{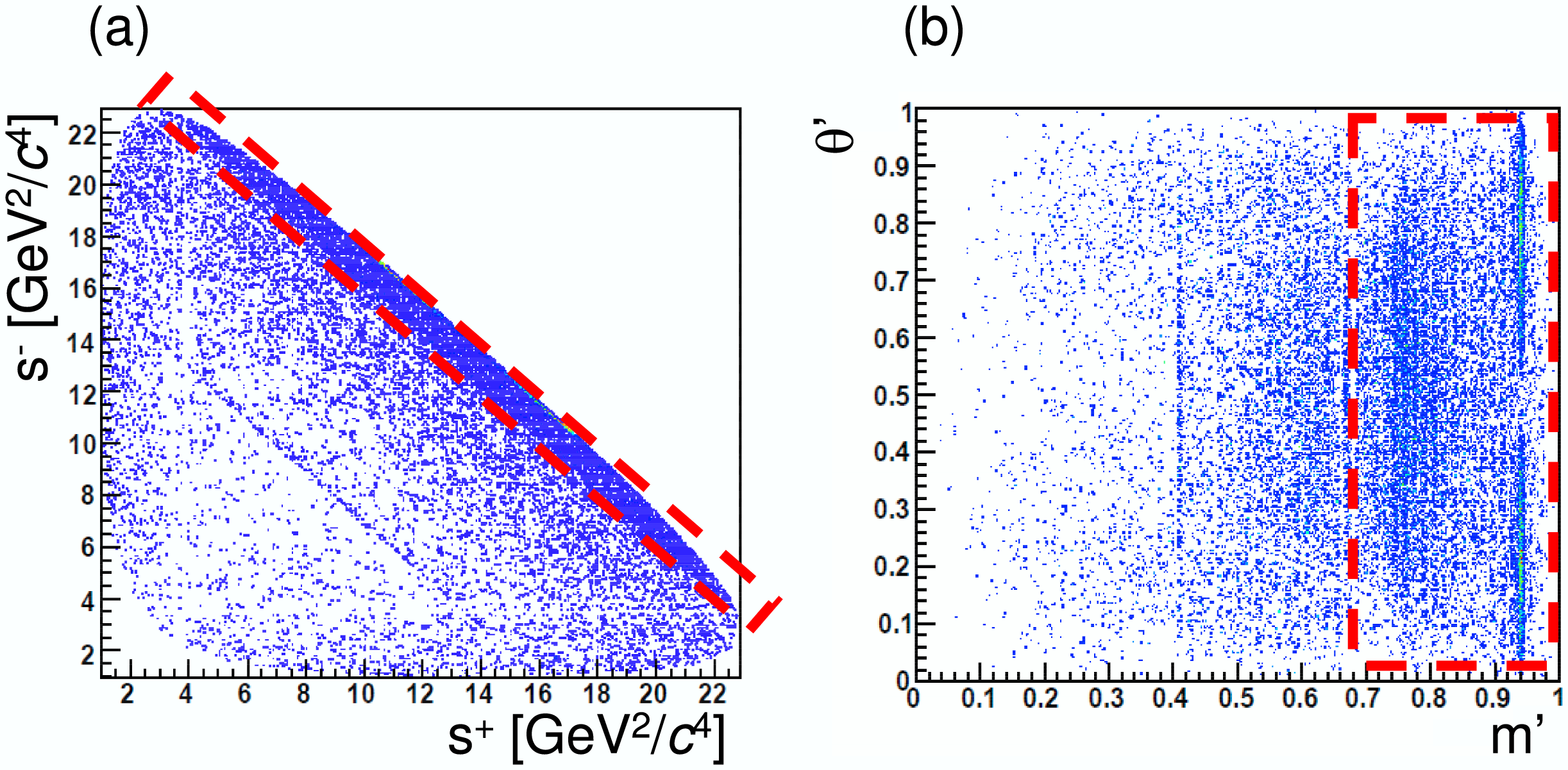}
 \caption{The Dalitz distribution based on our signal model of GEANT-based signal MC (a) with the normal Dalitz parameterization, $s_+$ and $s_-$, and (b) with the square Dalitz parameterization, $m'$ and $\theta '$. The dashed red boxes indicate the regions where most of the signal components and the background are located.}          
  \label{fig:usual_sq_dalitz}
 \end{center}
\rput[l](-4.3, 2.4) {\large{MC}}
\rput[l]( 3.0, 2.4) {\large{MC}}
\end{figure}

%${\bf the square Dalitz plot}$
In a Dalitz plot as a function of ($s_+$, $s_-$),
signal and continuum events densely populate the kinematic boundaries with low
$s_0$, which correspond to the $\phi(1020)$ and $f_0(980)$ resonances.
Large variations in a small area of the Dalitz plane make 
it difficult to use histograms to describe the background. 
Therefore, we apply the transformation,
\begin{equation}                                                            
ds_+ ds_- \to |{\rm det} J|dm'd\theta',
\end{equation}
where $J$ is the Jacobian of this transformation.
%($s_+$, $s_-$) $\to$ ($m'$, $\theta'$).
The parameters $m'$ and $\theta'$ are given by the transformation,
\begin{eqnarray}
m' &\equiv& {1 \over \pi}
\arccos \left(2 {{m_0 - m_0^{\rm min}}\over{m_0^{\rm max} - m_0^{\rm min}}}-1\right), \;{\rm and}   \label{eq:mp_def} \\
\theta' &\equiv& {1 \over \pi}\theta_0, \label{eq:tp_def}
\end{eqnarray}
where $m_0$ is $K^+K^-$ invariant mass, $m_0^{\rm max}$ and $m_0^{\rm min}$ are
kinematic limits of $m_0$, $\theta_0$ is the helicity angle,
defined as the angle between the $K^-$ and the $K^0_S$ in the $K^+ K^-$
rest frame.
With this transformation, the Dalitz plot turns into a ``square Dalitz plot''
with a smooth density variation.
Figure~\ref{fig:usual_sq_dalitz} (a) and (b) show the Dalitz distributions based on our signal model with the usual Dalitz parameterization, $s_+$ and $s_-$, the square Dalitz parameterization, $m'$ and $\theta '$, respectively.
As can be seen, the highlighted region where most of the signal and background events are located, is magnified in the square Dalitz parameterization.
The square Dalitz plot is described in detail
elsewhere~\cite{belle_pipiks_dalitz,babar_sdl_pipipi}.

%${\bf signal PDF}$
The PDF expected for the signal distribution, ${\cal P}_{\rm sig}$, is given by
\begin{equation}
{\cal P}_{\rm sig}(m', \theta', \Delta t, q) = 
\epsilon(m', \theta') | A_{\rm sig}(m', \theta', \Delta t, q) |^2
\otimes R_{\rm sig},
\end{equation}
where
\begin{eqnarray}
|A_{\rm sig}(m', \theta', \Delta t, q)|^2 
& = & |{\rm det} J| \frac{ e^{-|\Delta{t}|/{\tau_{B^0}}} }{4\tau_{B^0}}
 \Bigl[ (1-q \Delta w_{l})(|A|^2 + |\bar{A}|^2) \\ \nonumber
& - & q(1-2w_{l})(|A|^2 - |\bar{A}|^2)\cos(\Delta m_d \Delta{t}) \\ \nonumber
& + & 2q(1-2w_{l}){\mathcal Im}(\bar{A}A^*)\sin(\Delta m_d \Delta{t}) 
\Bigr],
\end{eqnarray}
which accounts for $CP$ dilution from the incorrect flavor tagging.
%$| A_{\rm sig} |^2$ is described by Eq.~\ref{signal_cp}, modified to
%incorporate the effect of incorrect flavor assignment:
This function is convolved with the $\Delta t $ resolution function
$R_{\rm sig}$~\cite{ref:belle_phiks_cp}; the impact of
detector resolution on the Dalitz plot is ignored because 
the intrinsic widths of the dominant resonances are
larger than the mass resolution.
We determine the variations of  the signal detection efficiency
across the Dalitz plane due to detector acceptance, 
$\epsilon(m', \theta')$,
by using a large MC sample.

%${\bf continuum PDF}$
The PDF for continuum background is
\begin{equation}
{\cal P}_{q\bar{q}}(m', \theta', \Delta t, q)
=\frac{1+q{\cal A}_{q\bar{q}}(\theta')}{2} 
H_{q\bar{q}}(m', \theta') P_{q\bar{q}}(\Delta t),
\end{equation}
where $H_{q\bar{q}}$, ${\cal A}_{q\bar{q}}$, and $P_{q\bar{q}}(\Delta t)$
are the Dalitz distribution PDF, the Dalitz-plot-dependent flavor asymmetry
and the $\Delta t$ PDF, respectively.
The function $P_{q\bar{q}}$ is modeled as a sum of exponential and prompt
components, and is convolved with a double Gaussian that represents the
resolution. All parameters of $P_{q\bar{q}}$ are determined by a fit to
%the $\Delta t$ distribution in the $\Delta E$-$M_{\rm bc}$ sideband region.
the $\Delta t$ distribution in the sideband region that is defined above.
The Dalitz distribution PDF, $H_{q\bar{q}}$, is a two-dimensional binned
histogram PDF.
To determine the PDF, we use the sideband region around the signal region
with a less restrictive requirement, $|\cos \theta_{\rm th}|<0.92$, to
increase statistics.
We have checked that the Dalitz distribution for this sideband region is similar to
that for the signal region, using a MC sample.
There is a flavor asymmetry ${\cal A}_{q\bar{q}}$
due to the jet-like topology of continuum
because a high momentum $K^+(K^-)$ in $f_{\rm rec}$ is accompanied
by a high momentum $K^-(K^+)$ in $f_{\rm tag}$; to
account for this, we extract the Dalitz plot asymmetry using
%a $\Delta E$-$M_{\rm bc}$ sideband region.
almost the same region as the region used in $H_{q\bar{q}}$ extraction:
since we find no correlation between $\theta '$ and $M_{\rm bc}$,
we enlarge the lower limit of the sideband region in this fit 
from $5.24$ GeV/$c^2$ to $5.2$ GeV/$c^2$ in $M_{\rm bc}$.

%${\bf charged and neutral B PDF}$
Using high-statistics MC sample, we find no $CP$ violating asymmetry in the
background coming from charmless and charmed $B$ decays.
Therefore, the PDFs for $B^+B^-$ and $B^0\overline{B}^0$ backgrounds are given by
\begin{eqnarray}
{\cal P}_{B^+B^-}(m', \theta', \Delta t) &=& H_{B^+B^-}(m', \theta')|A_{B^+B^-}(\Delta t)|^2\otimes R_{B^+B^-},\\
{\cal P}_{B^0\overline{B}^0}(m', \theta', \Delta t)&=&H_{B^0\overline{B}^0}(m', \theta')|A_{B^0\overline{B}^0}(\Delta t)|^2 \otimes R_{B^0\overline{B}^0},
\end{eqnarray}
respectively.
Dalitz distribution PDFs as well as $H_{B^+B^-(B^0\overline{B}^0)}$, are modeled with 
two-dimensional histograms from MC.
The $\Delta t$ PDF for both models, $|A_{B^+B^-(B^0\overline{B}^0)}|^2$,
are described by exponential functions with effective lifetimes while
$R_{B^+B^-(B^0\overline{B}^0)}$ 
are the $\Delta t$ resolution functions.
The effective lifetimes are obtained from fits to the MC sample.

%${\bf outlier PDF}$
To account for a small fraction of events with large $\Delta t$ values not yet described by either
signal or background PDFs, an outlier PDF is introduced,
${\cal P}_{ol}=H_{ol}G(\Delta t)$, where $G$ is a Gaussian and $H_{ol}$
is the two-dimensional binned histogram PDF of the Dalitz plot of data itself.

For the $j$-th event, the following likelihood function is evaluated:
\begin{eqnarray}
\label{total_pdf}
P_{j}(m', \theta', \Delta t, q; \Delta E,M_{\rm bc},l) &=& 
(1-f_{ol})\Biggl[\sum_{k}f_{k}(\Delta E,M_{\rm bc},l){\cal P}_k(m', \theta', \Delta t, q)\Biggr]
+f_{ol}{\cal P}_{ol}(m', \theta', \Delta t)
\end{eqnarray}
where $k$ runs over a total of four components including signal and backgrounds.
The probability of each component ($f_j$) is calculated using the
result of the $\Delta E$-$M_{\rm bc}$-$l$ fit on an event-by-event basis.
%For each contribution for signal, we can separately assign 
%the $CP$ violation parameters,

As there is only sensitivity to the relative amplitudes and phases
between decay modes, we fix $a_{(K^+K^-)_{\rm NR}}=60$ and $b_{(K^+K^-)_{\rm NR}}=0^{\circ}$.
In addition, $f_{\rm X}$ and non-resonant contributions are combined
and have a single common $CP$ violating parameters.
The combined component is referred to as ``others'' throughout this paper.
The parameters $d_{\chi_{c0}}$ and $c_{\chi_{c0}}$ are fixed to the world average
$b \to c\bar{c}s$ values of 21.5$^{\circ}$ and 0, respectively.
We determine 19 parameters of the Dalitz plot and $CP$ asymmetries by
maximizing the likelihood function ${\cal L}=\prod_j P_j$, where 
the product is over all events.

%${\bf fit result}$
We find four preferred solutions with consistent $CP$ parameters but significantly different
amplitudes for $f_0(980) K^0_S$ and $f_{X} K^0_S$.
The fitted results are summarized in Table~\ref{table:dtcp_results}.
These are obtained by performing a large number of fits with random input parameters.
For each resonance, $i$, the relative fractions can be calculated as
\begin{equation}
  f_{i} = \frac{(|a_{i}|^{2} + |\bar a_{i}|^{2}) \int F_{i}(s_{+}, s_{-}) F^{*}_{i}(s_{+}, s_{-}) ds_{+} ds_{-}}{\int  (|{A}|^{2} + |\bar {A}|^{2}) ds_{+} ds_{-}}, 
\end{equation}
where the sum of fractions over all decay channels may not be 100\% due
to interference.
Table~\ref{table:relative_frac} summarizes the relative fractions for all solutions.

%\begin{table}[htbp]
\begin{table}
\begin{center}
\caption{Time-dependent Dalitz plot fit results for the four solutions with statistical errors. The phases, $b_i$ and $d_i$, and $\alpha$ are given in degrees and GeV$^{-2}c^4$, respectively.}
\begin{tabular}{l|ccccc}
\hline\hline
Parameter & Solution 1 & Solution 2  & Solution 3 & Solution 4   \\ \hline 
%Dalitz Amplitude \\ 
$a_{f_0(980)}$ & 29.3 $_{-2.7}^{+2.6}$ & 53.0 $_{-19.9}^{+7.3}$ & 31.8 $_{-3.5}^{+3.0}$ & 64.1 $_{-5.8}^{+7.0}$ \\
$a_{\phi(1020)}$ & 0.53 $_{-0.06}^{+0.07}$ & 0.67 $_{-0.23}^{+0.10}$ & 0.56 $_{-0.06}^{+0.06}$ & 0.71 $_{-0.10}^{+0.13}$ \\
$a_{f_{\rm X}}$ & 5.2 $\pm$ 0.8 & 7.0 $_{-2.5}^{+1.2}$ & 15.6 $_{-1.4}^{+1.5}$ & 23.9 $_{-3.1}^{+3.9}$ \\
$a_{\chi_{\rm c0}}$ & 2.03 $_{-0.28}^{+0.31}$ & 2.53 $_{-0.89}^{+0.46}$ & 2.16 $_{-0.28}^{+0.29}$ & 2.89 $_{-0.45}^{+0.56}$ \\
$a_{(K^0_SK^+)_{\rm NR}}$ & 6.5 $_{-6.7}^{+8.2}$ & 20.8 $_{-7.6}^{+9.0}$ & 10.3 $_{-6.2}^{+7.5}$ & 24.3 $_{-6.2}^{+6.8}$ \\
$a_{(K^0_SK^-)_{\rm NR}}$ & 25.9 $_{-4.3}^{+4.9}$ & 40.2 $_{-6.4}^{+7.1}$ & 29.7 $_{-5.4}^{+6.4}$ & 21.7 $_{-6.0}^{+5.9}$ \\  \hline
%Dalitz Phase (deg.) \\
$b_{f_0(980)}$ & -16.0 $_{-13.2}^{+10.0}$ & 83.4 $_{-7.3}^{+8.8}$ & -1.8 $_{-13.5}^{+10.0}$ & 109.3 $_{-8.2}^{+10.5}$ \\
$b_{\phi(1020)}$ & -34.5 $_{-14.5}^{+14.0}$ & 108.7 $_{-15.5}^{+15.8}$ & -7.0 $_{-14.5}^{+14.1}$ & 149.2 $_{-16.6}^{+17.5}$ \\
$b_{f_{\rm X}}$ & -32.6 $_{-8.9}^{+8.3}$ & -106.0 $_{-13.8}^{+12.9}$ & 92.1 $_{-8.2}^{+8.7}$ & 35.7 $_{-5.9}^{+6.9}$ \\
$b_{\chi_{\rm c0}}$ & -28.0 $_{-26.6}^{+22.5}$ & -36.2 $_{-28.3}^{+25.0}$ & -35.7 $_{-27.8}^{+25.3}$ & 45.96 $_{-30.2}^{+22.4}$ \\
$b_{(K^0_SK^+)_{\rm NR}}$ & 126.5 $_{-82.7}^{+39.7}$ & 113.6 $_{-16.2}^{+13.6}$ & 113.5 $_{-31.4}^{+19.0}$ & 98.8 $_{-21.8}^{+23.0}$ \\
$b_{(K^0_SK^-)_{\rm NR}}$ & -123.5 $_{-11.4}^{+13.4}$ & -143.5 $_{-7.6}^{+8.3}$ & -141.2 $_{-8.8}^{+9.8}$ & 23.7 $_{-24.6}^{+22.2}$  \\
%$b_{\rm NR^0}$ & 0.0(fixed) & 0.0(fixed) & 0.0(fixed) & 0.0(fixed)  \\
 \hline
$c_{f_0(980)}$ & 0.16 $_{-0.15}^{+0.16}$ & 0.10 $_{-0.07}^{+0.08}$ & -0.01 $\pm$ {0.11} & 0.09 $\pm$ {0.07} \\
$c_{\phi(1020)}$ & -0.02 $\pm$ {0.10} & -0.04 $\pm$ {0.09} & 0.01 $\pm$ {0.10} & -0.10 $\pm$ {0.09} \\
$c_{\rm others}$ & 0.07 $\pm$ {0.06} & 0.03 $\pm$ {0.08} & 0.01 $_{-0.05}^{+0.04}$ & -0.02 $_{-0.06}^{+0.05}$ \\  \hline
%$\phi_{1}$ (deg.) \\ 
$d_{f_0(980)}$ & 31.3 $_{-8.5}^{+9.0}$ & 26.1 $_{-6.6}^{+7.0}$ & 25.6 $_{-7.2}^{+7.6}$ & 26.3 $_{-5.4}^{+5.7}$ \\
$d_{\phi(1020)}$ & 32.2 $_{-8.4}^{+9.0}$ & 26.2 $_{-8.4}^{+8.8}$ & 27.3 $_{-8.0}^{+8.6}$ & 24.3 $_{-7.7}^{+8.0}$ \\
$d_{\rm others}$ & 24.9 $_{-6.0}^{+6.4}$ & 29.8 $_{-6.4}^{+6.6}$ & 26.2 $_{-5.4}^{+5.9}$ & 23.8 $_{-5.1}^{+5.5}$ \\
\hline 
$\alpha$   & 0.12 $_{-0.04}^{+0.03}$ & 0.06 $\pm$ {0.04} & 0.10 $\pm$ {0.04} & 0.18 $\pm$ {0.03} \\
  \hline
$-2\log{\cal L}$ & 10201.7 & 10198.6 & 10204.5 & 10208.9  \\
 \hline \hline
\end{tabular}
\label{table:dtcp_results}
\end{center}
\end{table}

%\begin{table}[hbtp]
\begin{table}
\begin{center}
\caption{
Summary of the relative fractions (\%), the errors are statistical only.
}
\begin{tabular}{l|ccccc}
 \hline\hline
Parameter & Solution 1 & Solution 2  & Solution 3 & Solution 4\\ \hline
%$-2\Delta \log{\cal L}$ & 3.135 (2nd) & 0 (best) & 5.912 (3rd) & 10.310 (4th)  \\ \hline 
%Relative fraction ($\%$) \\
$f_{f_0(980) K^0_S}$ & 26.0 $\pm$ 7.4 & 54.0 $\pm$ 9.6 & 26.4 $\pm$ 7.8 & 68.1 $\pm$ 12.3 \\
$f_{\phi(1020) K^0_S}$ & 14.2 $\pm$ 1.2 & 14.5 $\pm$ 1.2 & 14.2 $\pm$ 1.2 & 14.4 $\pm$ 1.2 \\
$f_{f_{\rm X} K^0_S}$ & 5.10 $\pm$ 1.39 & 5.89 $\pm$ 1.86 & 39.6 $\pm$ 2.6 & 59.0 $\pm$ 3.0 \\
$f_{\chi_{c0} K^0_S}$ & 3.73 $\pm$ 0.74 & 3.71 $\pm$ 0.73 & 3.68 $\pm$ 0.73 & 4.15 $\pm$ 0.79 \\
$f_{(K^+K^-)_{\rm NR}K^0_S}$ & 138.4 $\pm$ 44.8 & 175.0 $\pm$ 52.6 & 157.4 $\pm$ 29.5 & 48.1 $\pm$ 11.7 \\
$f_{(K^0_SK^+)_{\rm NR}K^-}$ & 1.65 $\pm$ 4.17 & 21.0 $\pm$ 17.3 & 4.63 $\pm$ 6.76 & 7.87 $\pm$ 4.78 \\
$f_{(K^0_SK^-)_{\rm NR}K^+}$ & 26.0 $\pm$ 12.9 & 78.0 $\pm$ 36.2 & 38.6 $\pm$ 18.1 & 6.27 $\pm$ 3.81 \\
\hline
$F_{\rm tot}$ & 215.2 $\pm$ 47.5 & 352.0 $\pm$ 66.8 & 284.5 $\pm$ 36.3 & 207.9 $\pm$ 18.4 \\
\hline  \hline
\end{tabular}
\label{table:relative_frac}
\end{center}
\end{table}

By translating the fit results using Eqs.~\ref{eq:acp_dtcpv} and~\ref{eq:phione_dtcpv}, 
we determine the time-dependent $CP$ violating parameters
of $B^0 \to f_0(980) K^0_S$ and $B^0 \to \phi(1020) K^0_S$ decays and
other $B^0$ decays with the $K^+K^-K^0_S$ final state.
Table~\ref{table:tcpv_results} summarizes the $CP$ violating parameters for all solutions.

%\begin{table}[hbtp]
\begin{table}
\begin{center}
\caption{Time-dependent $CP$ violating parameters for the four solutions, where the first error is statistical, the second is systematic and the third is the Dalitz plot model uncertainty.
}
\begin{tabular}{l|cccc}\hline\hline
& Solution 1 & Solution 2 & Solution 3 & Solution 4  \\ \hline
% A_fz
${\cal A}_{CP}(f_0(980) K^0_S)$ & $-0.30 \pm 0.29 \pm 0.11 \pm 0.09$
& $-0.20 \pm 0.15  \pm 0.08 \pm 0.05$
& $+0.02 \pm 0.21  \pm 0.09 \pm 0.09$
& $-0.18 \pm 0.14  \pm 0.08 \pm 0.06$  \\
% phi1_fz
$\phi_1^{\rm eff}(f_0(980) K^0_S)$ & $(31.3 \pm 9.0 \pm 3.4 \pm 4.0)^{\circ}$
& $(26.1 \pm 7.0 \pm 2.4 \pm 2.5 )^{\circ}$
& $(25.6 \pm 7.6 \pm 2.9 \pm 0.8 )^{\circ}$
& $(26.3 \pm 5.7 \pm 2.4 \pm 5.8 )^{\circ}$ \\
% A_phi
${\cal A}_{CP}(\phi(1020) K^0_S)$ & $+0.04 \pm 0.20 \pm 0.10 \pm 0.02$
& $+0.08 \pm 0.18  \pm 0.10 \pm 0.03$
& $-0.01 \pm 0.20  \pm 0.11 \pm 0.02$
& $+0.21 \pm 0.18 \pm 0.11 \pm 0.05 $ \\
% phi1_phi
$\phi_1^{\rm eff}(\phi(1020) K^0_S)$ & $(32.2 \pm 9.0 \pm 2.6 \pm 1.4)^{\circ}$
& $(26.2 \pm 8.8 \pm 2.7 \pm 1.2 )^{\circ}$
& $(27.3 \pm 8.6 \pm 2.8 \pm 1.3 )^{\circ}$
& $(24.3 \pm 8.0 \pm 2.9 \pm 5.2 )^{\circ}$ \\
% A_NR
${\cal A}_{CP}({\rm others})$ & $-0.14 \pm 0.11 \pm 0.08 \pm 0.03$
& $-0.06 \pm 0.15 \pm 0.08 \pm 0.04$
& $-0.03 \pm 0.09 \pm 0.08 \pm 0.03$
& $+0.04 \pm 0.11 \pm 0.08 \pm 0.02$ \\
% NR 0
$\phi_1^{\rm eff}({\rm others})$ & $(24.9 \pm 6.4 \pm 2.1 \pm 2.5)^{\circ}$
& $(29.8 \pm 6.6 \pm 2.1 \pm 1.1)^{\circ}$
& $(26.2 \pm 5.9 \pm 2.3 \pm 1.5)^{\circ}$
& $(23.8 \pm 5.5 \pm 1.9 \pm 6.4)^{\circ}$ \\
\hline\hline
\end{tabular}
\label{table:tcpv_results}
\end{center}
\end{table}
%Based on the toyMC study and Likelihood check, 
%Solution1 is the most preferable and is considered as our result.

\begin{figure}
\begin{center}
\includegraphics[width=0.45\textwidth]{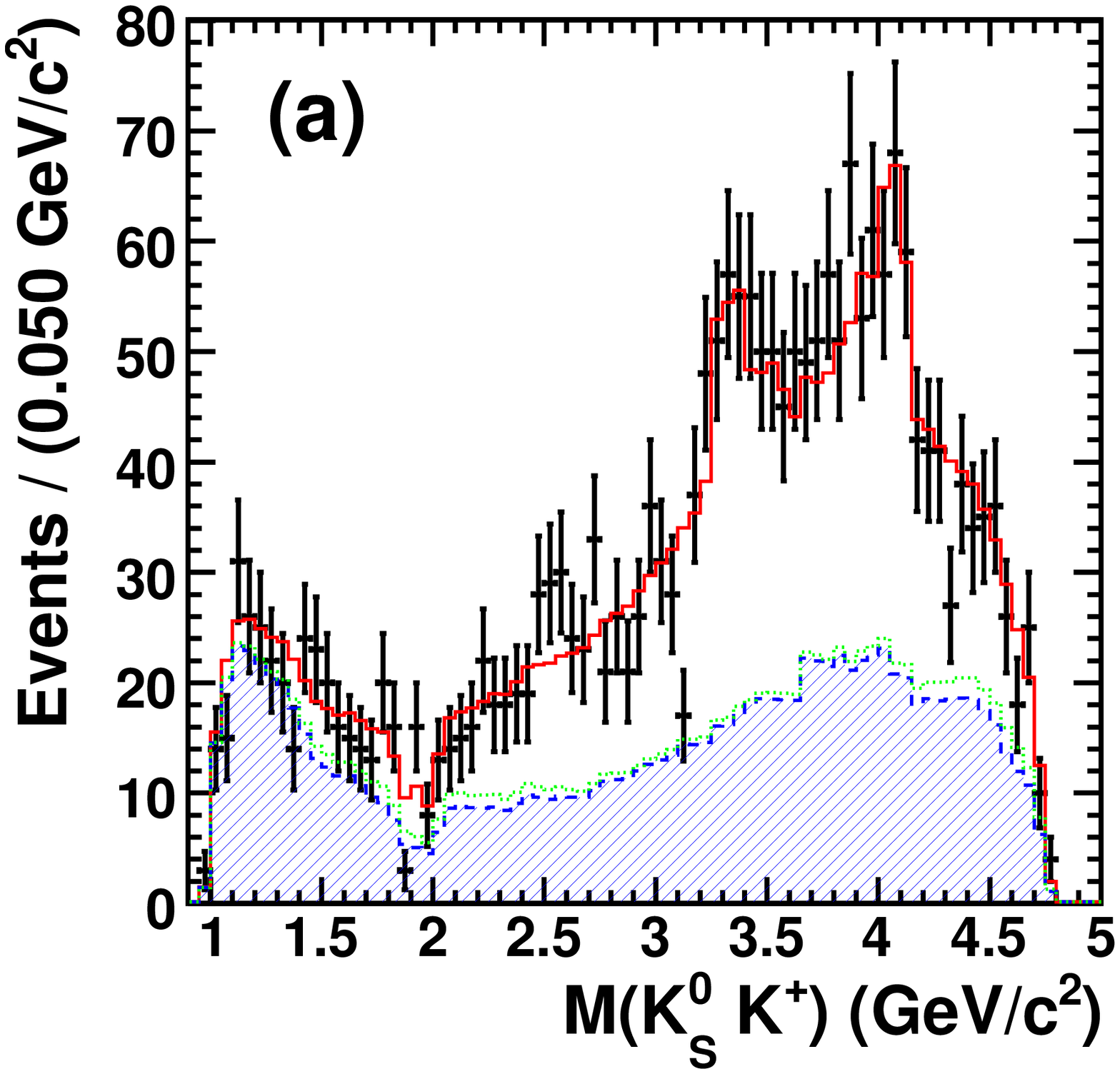}
\includegraphics[width=0.45\textwidth]{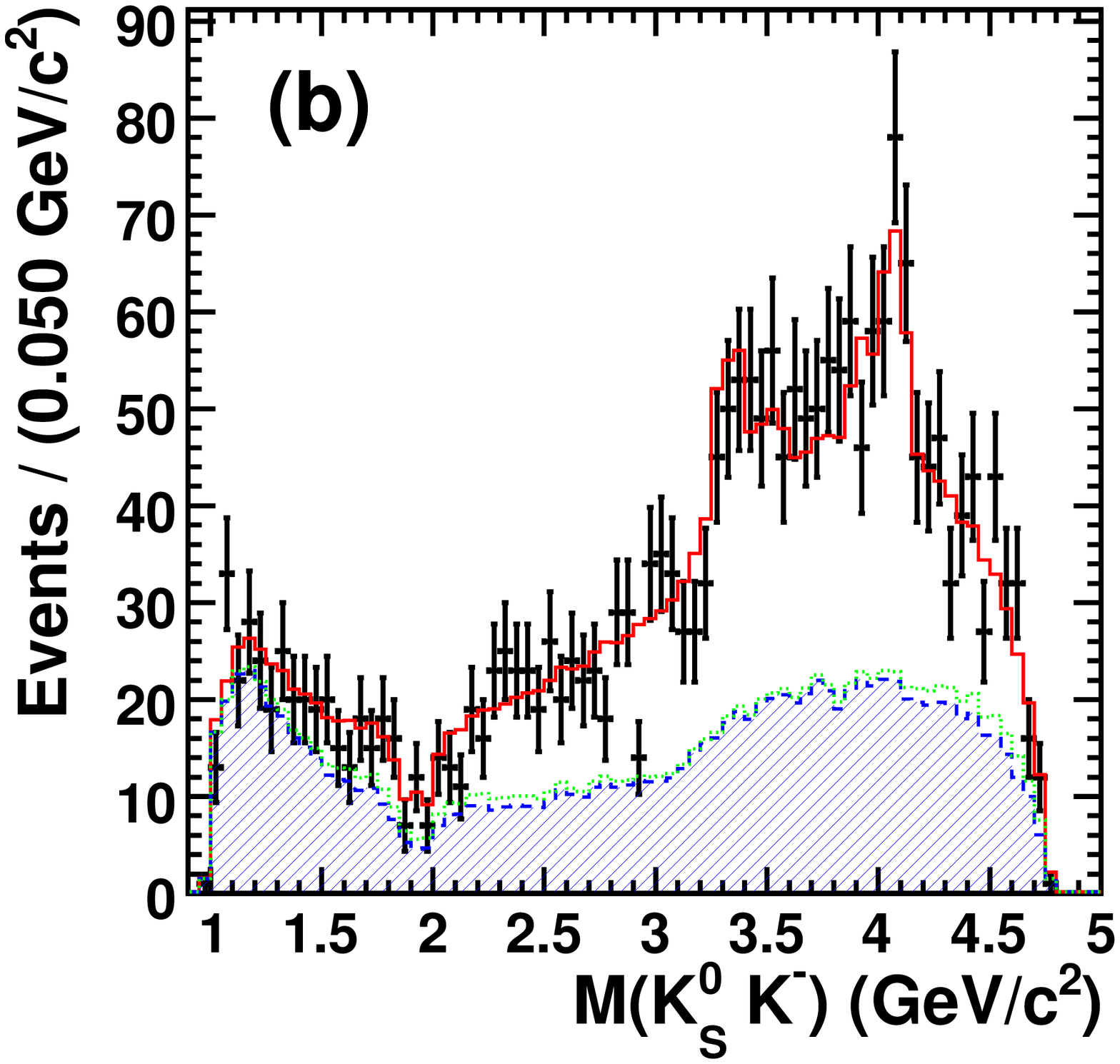}
\includegraphics[width=0.45\textwidth]{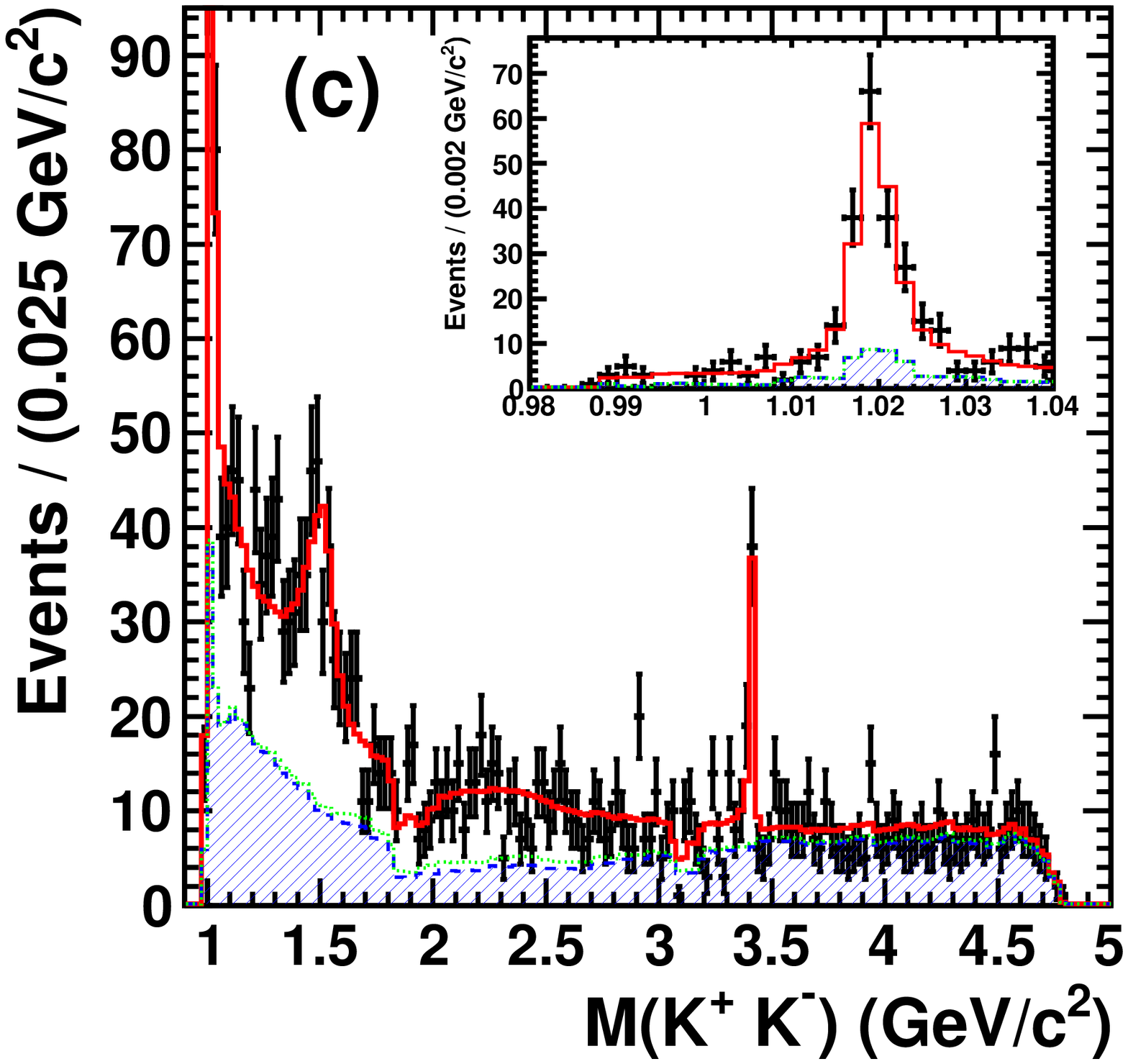}

\caption{The mass projections onto (a) $M(K^0_S K^+)$, (b) $M(K^0_S K^-)$ and (c) $M(K^+ K^-)$ distributions (the inset shows the projection near the $\phi(1020)$ resonance) for the $B^0 \rightarrow K^+K^-K^0_S$ candidate events in the signal region, using Solution 1. In (a-c), the red solid curves show the fit projections while the blue hatched areas and the green dashed curves show the $q\bar{q}$ and total background components, respectively. The points with error bars are the data.}
\label{fig:kkks_dl}
\end{center}
\end{figure}

\begin{figure}
\begin{center}
\includegraphics[width=0.45\textwidth]{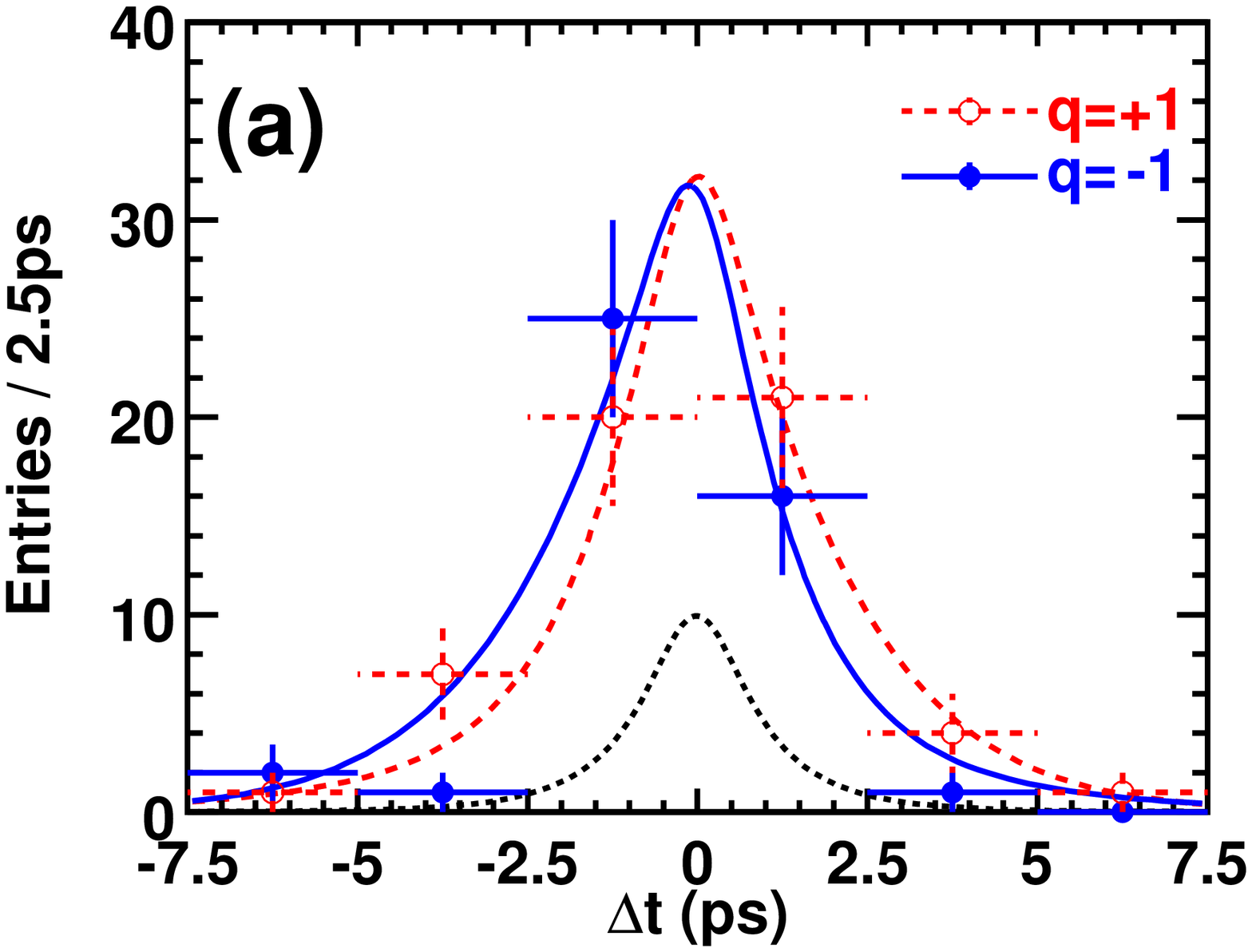}
\includegraphics[width=0.45\textwidth]{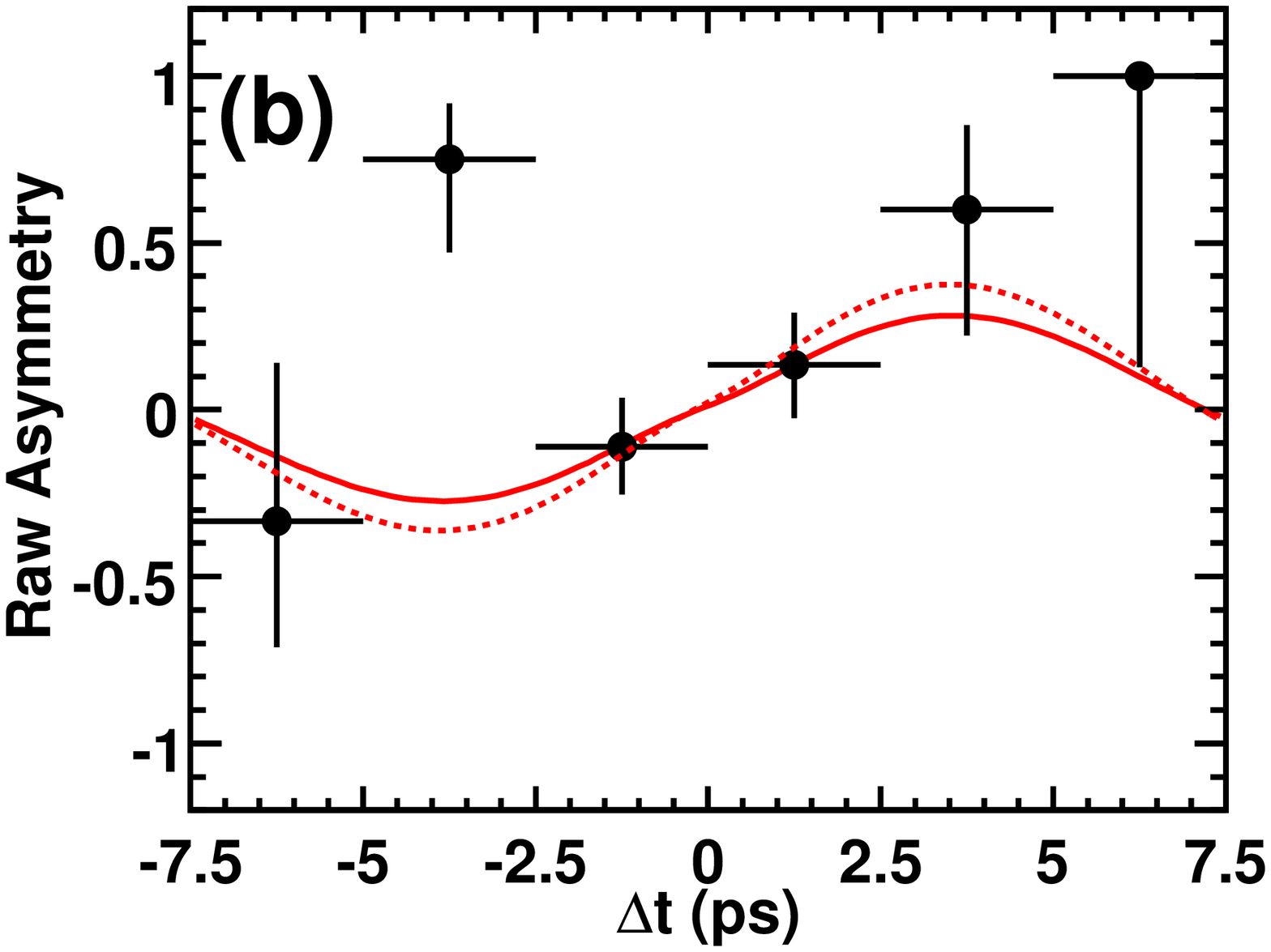}
\caption{
(a) $\Delta t$ distribution and (b) raw asymmetry for the $B^0 \rightarrow K^+K^-K^0_S$ candidates in the $\phi(1020) K^0_S$ region, $|M_{K^+K^-}-M_{\phi(1020)}|<0.01$ GeV/$c^2$, with good tags, $r > 0.5$, using Solution 1. In (a), the blue solid and red dashed curves show the fitted results with $B^0$ and $\overline{B}^0$ tags, respectively. The dotted black curve shows the background component with $B^0$ and $\overline{B}^0$ tags.
In (b), the solid curve shows the fit projection and the dashed curve shows the Standard Model expectation from the time-dependent $CP$ asymmetry measurement in $b \to c{\overline c}s$ decays.}
\label{fig:kkks_dt}
\end{center}
\end{figure}

\begin{figure}
\begin{center}
\includegraphics[width=0.45\textwidth]{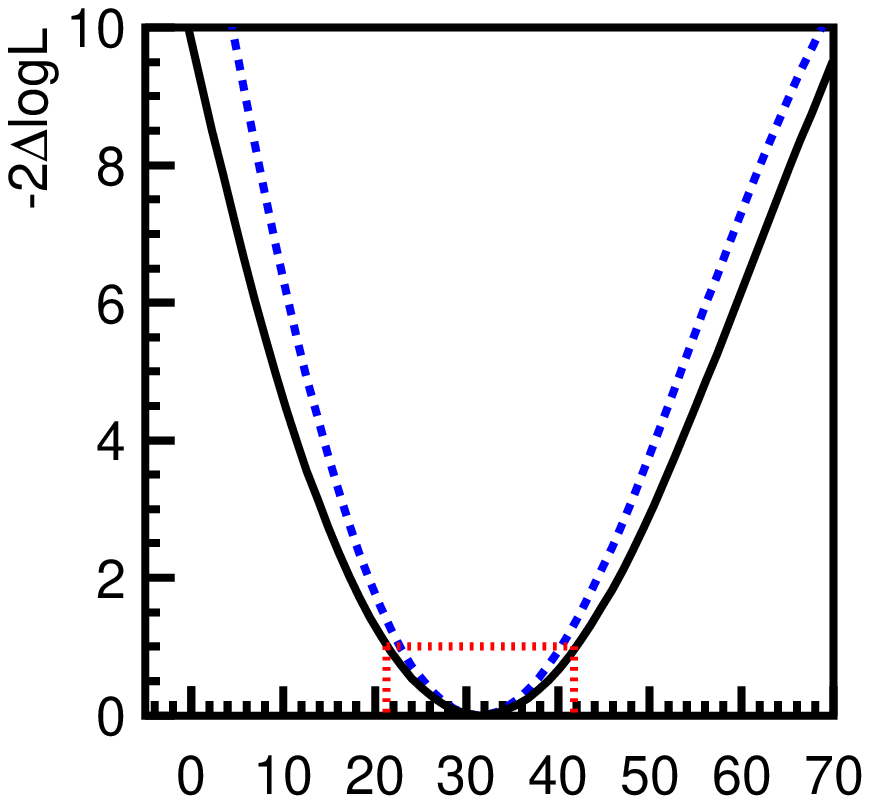}
\includegraphics[width=0.45\textwidth]{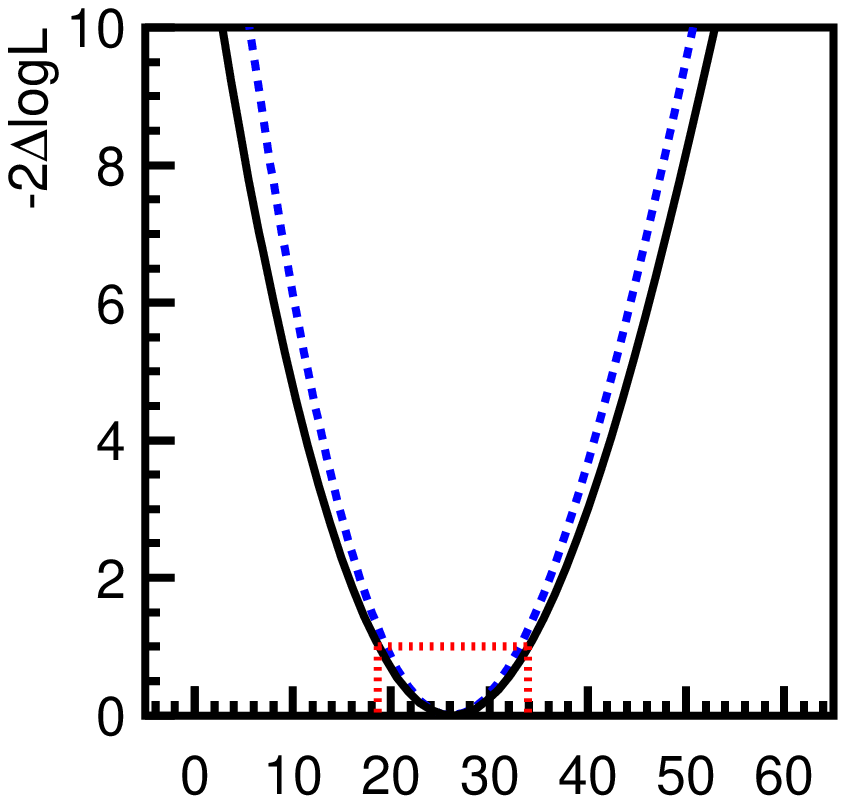}
\includegraphics[width=0.45\textwidth]{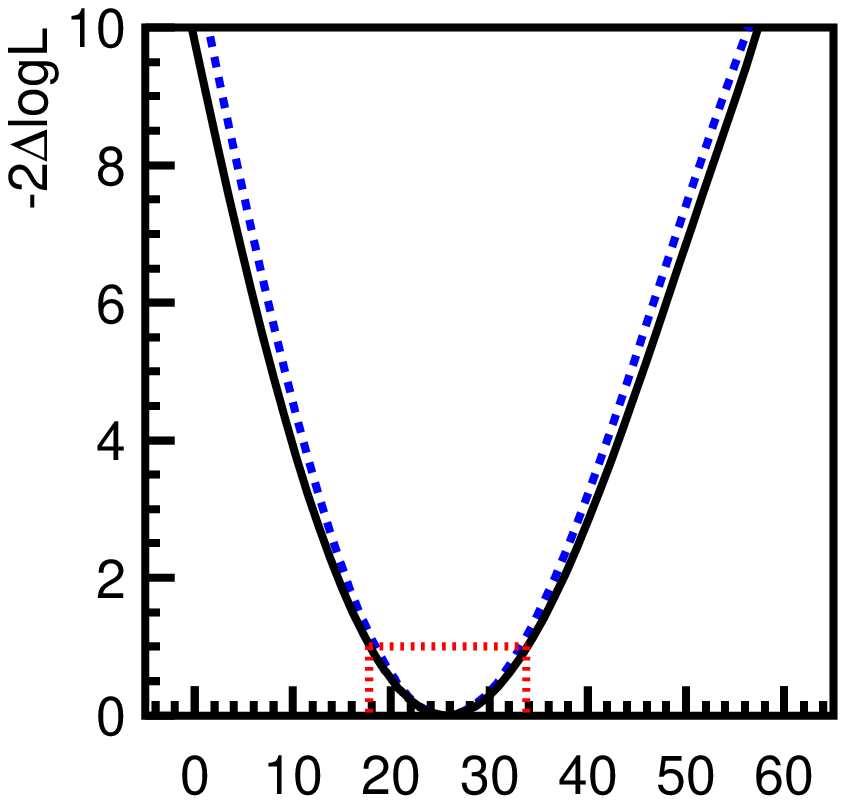}
\includegraphics[width=0.45\textwidth]{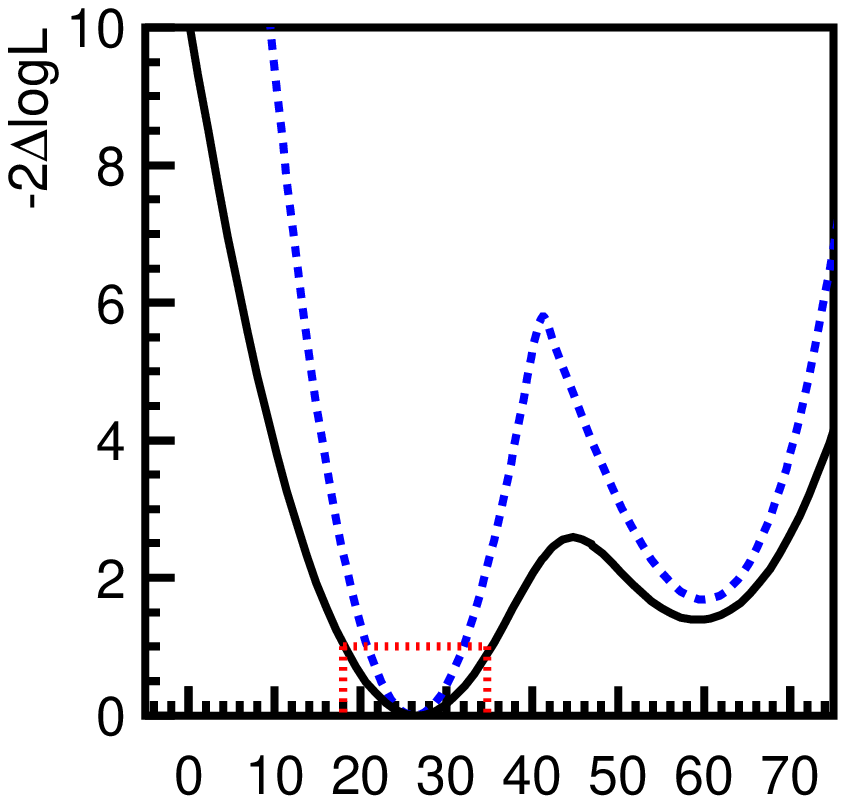}
\caption{
Likelihood scans of $\phi_1^{\rm eff}(f_0(980) K^0_S)$ for (a) Solution 1,
(b) Solution 2, (c) Solution 3, and (d) Solution 4.
The solid (dashed) curve contains the total (statistical) error
and the dotted box indicates the parameter range corresponding to $\pm 1\sigma$.
}
\label{fig:scan_f0ks}
\end{center}
\rput[l](-5.5, 16.7) {\Large{(a)}}
\rput[l]( 3. , 16.7) {\Large{(b)}}
\rput[l](-5.5,  8.7) {\Large{(c)}}
\rput[l]( 3. ,  8.7) {\Large{(d)}}
\rput[l](-4.5, 10.4) {\large{$\phi_1^{\rm eff}(f_0(980) K^0_S)$}}
\rput[l]( 4. , 10.4) {\large{$\phi_1^{\rm eff}(f_0(980) K^0_S)$}}
\rput[l](-4.5,  2.4) {\large{$\phi_1^{\rm eff}(f_0(980) K^0_S)$}}
\rput[l]( 4. ,  2.4) {\large{$\phi_1^{\rm eff}(f_0(980) K^0_S)$}}
\end{figure}

\begin{figure}
\begin{center}
\includegraphics[width=0.45\textwidth]{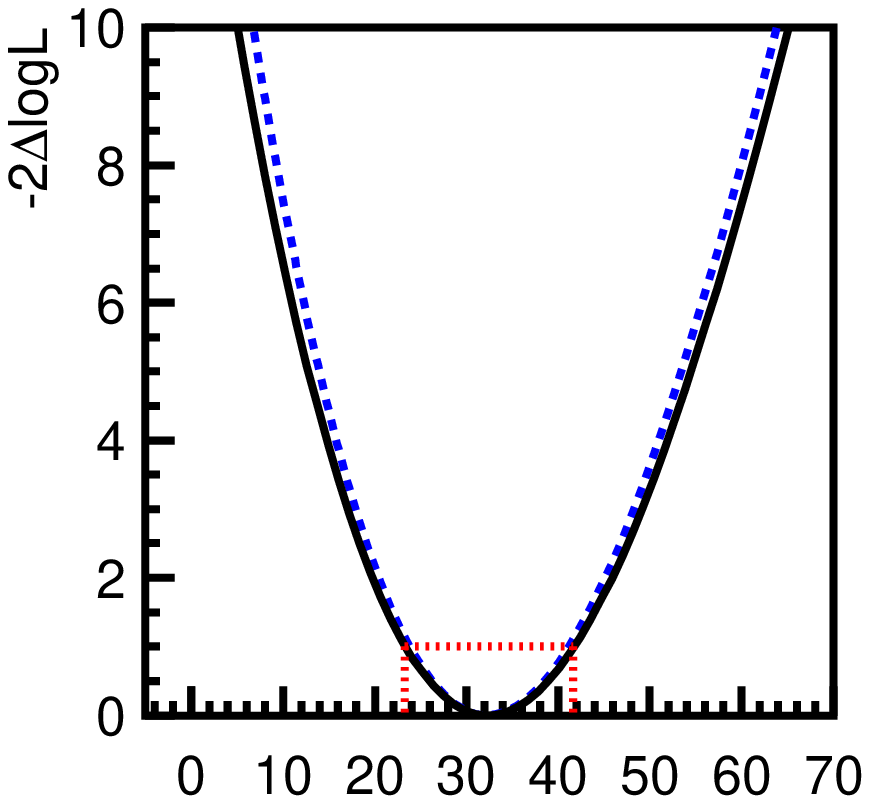}
\includegraphics[width=0.45\textwidth]{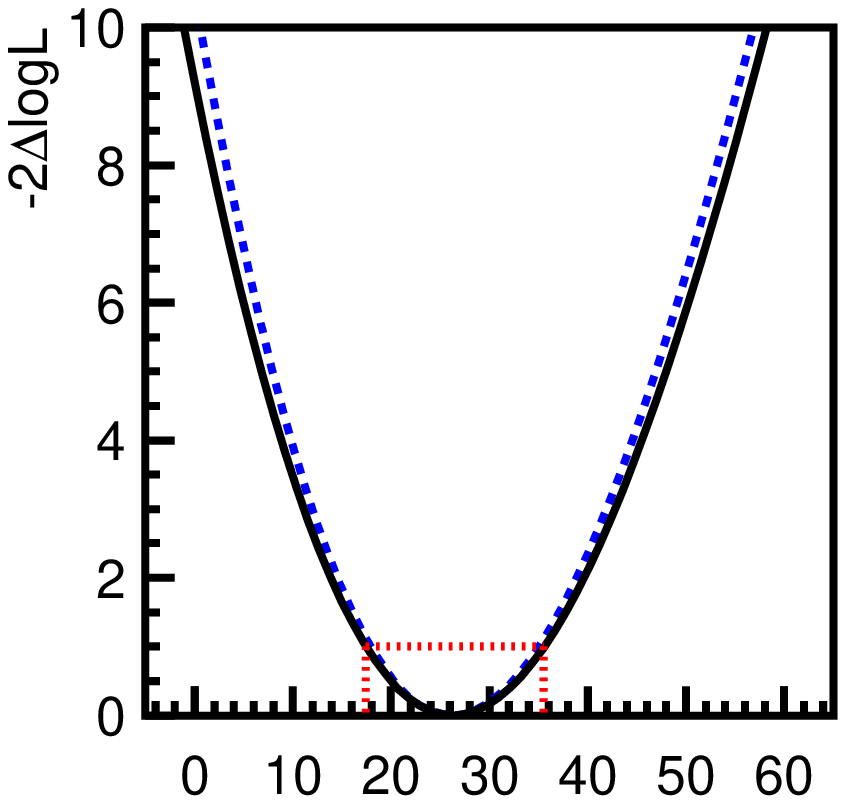}
\includegraphics[width=0.45\textwidth]{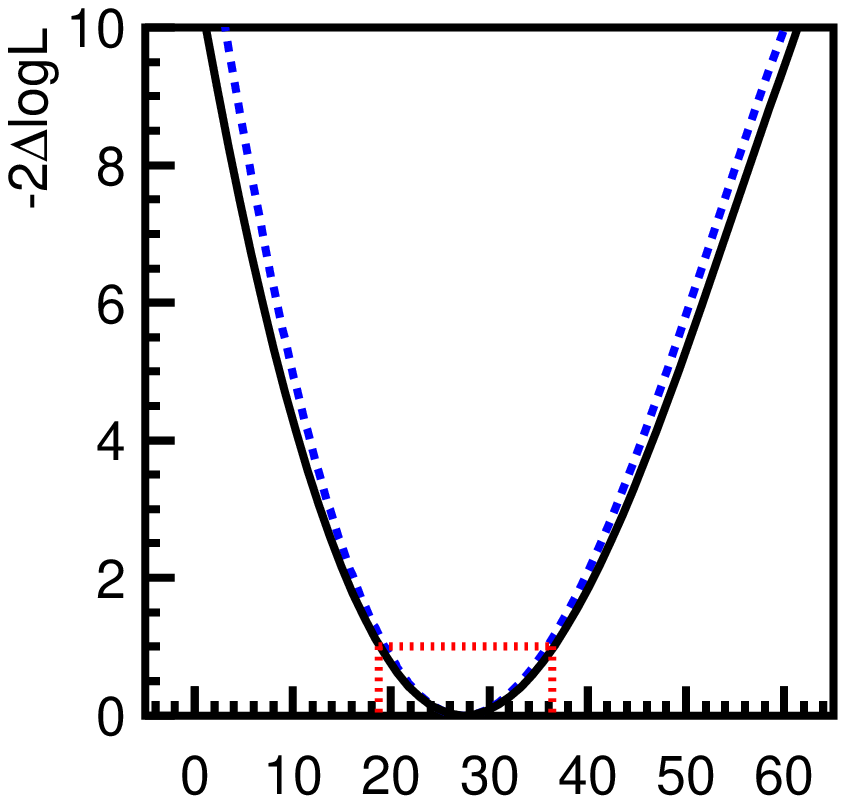}
\includegraphics[width=0.45\textwidth]{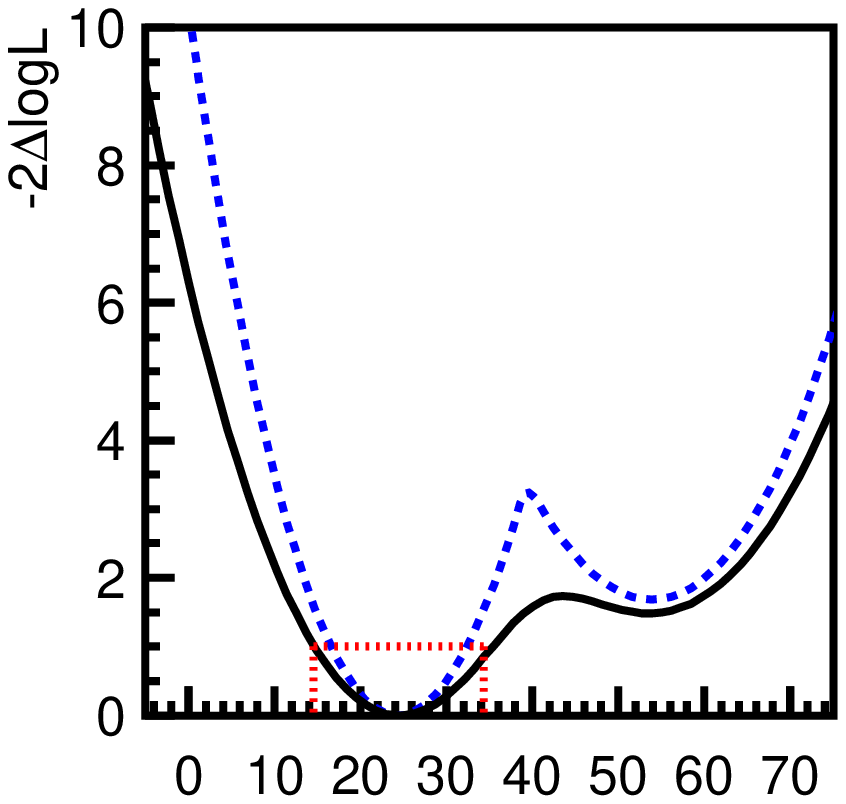}
\caption{
Likelihood scans of $\phi_1^{\rm eff}(\phi(1020) K^0_S)$ for (a) Solution 1,
(b) Solution 2, (c) Solution 3, and (d) Solution 4.
The solid (dashed) curve contains the total (statistical) error
and the dotted box indicates the parameter range corresponding to $\pm 1\sigma$.
}
\label{fig:scan_phiks}
\end{center}
\rput[l](-5.5, 16.7) {\Large{(a)}}
\rput[l]( 3. , 16.7) {\Large{(b)}}
\rput[l](-5.5,  8.7) {\Large{(c)}}
\rput[l]( 3. ,  8.7) {\Large{(d)}}
\rput[l](-4.5, 10.4) {\large{$\phi_1^{\rm eff}(\phi(1020) K^0_S)$}}
\rput[l]( 4. , 10.4) {\large{$\phi_1^{\rm eff}(\phi(1020) K^0_S)$}}
\rput[l](-4.5,  2.4) {\large{$\phi_1^{\rm eff}(\phi(1020) K^0_S)$}}
\rput[l]( 4. ,  2.4) {\large{$\phi_1^{\rm eff}(\phi(1020) K^0_S)$}}
\end{figure}

%${\bf Selection of the Most Preferable Solution Using External Information}$
In Table~\ref{table:relative_frac}, $f_{\phi(1020) K^0_S}$ is similar
for all four solutions but $f_{f_0(980) K^0_S}$ and $f_{f_{\rm X} K^0_S}$
are significantly different.
These four solutions are due to interference between the $f_0(980)$ and
non-resonant component, and interference between the $f_{\rm X}$ and non-resonant component,
and are characterized by different relative fractions for $f_0(980)$ and $f_{\rm X}$.
In order to distinguish these solutions with the current statistics, we
use external information from $B^0 \to \pi^+ \pi^- K^0_S$ and the
property that $f_0(980)$ decays to either $\pi^+\pi^-$ or $K^+ K^-$. 
We calculate the branching fraction, ${\cal B} (B^0 \to f_0(980)[ \pi^+ \pi^-] K^0_S)$,
based on the branching fraction of $B^0 \to \pi^+ \pi^- K^0_S$
and the relative fraction of $f_0(980) K^0_S$ in the $B^0 \to \pi^+ \pi^- K^0_S$ decay~\cite{garmash_kspipi}.
Similarly, we can calculate the branching fraction, ${\cal B} (B^0 \to f_0(980)[ K^+ K^-] K^0_S)$ from
Table~\ref{table:relative_frac}.
The fraction,
$ f_{f_0(980) \to \pi \pi }=\frac{{\cal B} (f_0(980) \to \pi  \pi)}{{\cal B} (f_0(980) \to \pi \pi) + {\cal B} (f_0(980) \to K K)}$,
is calculated to be $0.47 \pm 0.10$ for Solution 1 and $0.30 \pm 0.07$
for Solution 2.
The value of $f_{f_0(980) \to \pi \pi }$ is also determined by the BES
Collaboration, which uses the same parameterization for the $f_0(980)$, to be
$0.75 \pm 0.12$~\cite{ref:flattetwo}.
Therefore, the solutions with a low $f_0(980)K^0_S$ fraction (Solution 1 and 3) are preferred.
It is likely that the $f_{\rm X}$ described in this analysis and the $B^0 \to \pi^+ \pi^- K^0_S$ analysis~\cite{garmash_kspipi},
is the same state, $f_0 (1500)$.
If this is the case, the ratio,
$ \frac{{\cal B}(f_0 (1500) \to \pi \pi)}{{\cal B}(f_0 (1500) \to K K)}$,
is calculated to be $1.0 \pm 0.7$ for Solution 1 and $0.13 \pm 0.09$ for
Solution 3. As the world average of this ratio is $4.1 \pm 2.5$~\cite{ref:pdg},
the solutions with a low $f_{\rm X} K^0_S$ fraction (Solution 1 and 2) 
are preferred.
Table~\ref{table:comp_sol} summarizes these values for each solution.
Altogether, we conclude that Solution 1 is preferred
from all currently available external measurements.
The mass projections onto (a) $M(K^0_S K^+)$, (b) $M(K^0_S K^-)$ and (c) $M(K^+ K^-)$ distributions for Solution 1 are shown in Fig.~\ref{fig:kkks_dl},
and (a) $\Delta t$ distribution and (b) raw asymmetry in the $\phi(1020) K^0_S$ region are shown in Fig.~\ref{fig:kkks_dt}.
The full correlation matrix is given is Tables~\ref{table:full_cor_matrix1_1} and ~\ref{table:full_cor_matrix1_2}.
Likelihood scans of $\phi_1^{\rm eff}$ for all four solutions are obtained
by fixing $\phi_1^{\rm eff}$ and redoing the fit.
We also perform scans that include the systematic and model errors by convolving the likelihood with a Gaussian with the width set to the quadratic sum of the systematic and mode uncertainties.
These are shown in Figs.~\ref{fig:scan_f0ks},~\ref{fig:scan_phiks}.

\begin{table}
\begin{center}
\caption{Comparison of external information with each of the four solutions.}
\begin{tabular}{l|ll}\hline\hline
 & $f_{f_0(980) \to \pi \pi }$ & $ \frac{{\cal B}(f_0 (1500) \to \pi \pi)}{{\cal B}(f_0 (1500) \to K K)}$ \\ \hline
Solution 1 & $0.47 \pm 0.10$ & $1.0 \pm 0.7$ \\
Solution 2 & $0.30 \pm 0.07$ & $0.91 \pm 0.64$ \\
Solution 3 & $0.46 \pm 0.10$ & $0.13 \pm 0.09$ \\
Solution 4 & $0.25 \pm 0.06$ & $0.09 \pm 0.06$ \\ \hline
External information & $0.75 \pm 0.12$~\cite{ref:flattetwo} & $4.1 \pm 2.5$~\cite{ref:pdg} \\
\hline\hline
\end{tabular}
\label{table:comp_sol}
\end{center}
\end{table}

%\begin{figure}[hbtp]

%\begin{figure}[htbp]

%\begin{table}[htbp]
\begin{table}
\begin{center}
\caption{Statistical correlation matrix for Solution1.}
\begin{tabular}{l|ccccc ccccc} 
\hline \hline
& $a_{f_0(980)}$ & $a_{\phi(1020) K^0_S}$ & $a_{f_{X} K^0_S}$ & $a_{\chi_{c0}K^0_S}$  & $a_{(K^0_SK^+)_{\rm NR}}$ & $a_{(K^0_SK^-)_{\rm NR}}$ & $b_{f_0(980)}$ & $b_{\phi(1020) K^0_S}$ & $b_{f_{X} K^0_S}$ & $b_{\chi_{c0}K^0_S}$ \\ \hline
$a_{f_0(980)}$            &1.00 & 0.13 & 0.42 & 0.13 & 0.29 & 0.17 &-0.04 & 0.22 &-0.18 & 0.14 \\
$a_{\phi(1020) K^0_S}$    &     & 1.00 & 0.57 & 0.71 & 0.54 & 0.71 &-0.88 &-0.35 &-0.38 & 0.15 \\
$a_{f_{X} K^0_S}$         &     &      & 1.00 & 0.45 & 0.50 & 0.51 &-0.44 &-0.09 &-0.09 & 0.12 \\
$a_{\chi_{c0}K^0_S}$      &     &      &      & 1.00 & 0.41 & 0.52 &-0.71 &-0.28 &-0.31 & 0.37 \\
$a_{(K^0_SK^+)_{\rm NR}}$ &     &      &      &      & 1.00 & 0.67 &-0.51 &-0.05 &-0.12 & 0.17 \\
$a_{(K^0_SK^-)_{\rm NR}}$ &     &      &      &      &      & 1.00 &-0.73 &-0.14 &-0.29 &-0.01 \\
$b_{f_0(980)}$            &     &      &      &      &      &      & 1.00 & 0.38 & 0.40 &-0.09 \\
$b_{\phi(1020) K^0_S}$    &     &      &      &      &      &      &      & 1.00 & 0.17 &-0.01 \\
$b_{f_{X} K^0_S}$         &     &      &      &      &      &      &      &      & 1.00 &-0.01 \\
$b_{\chi_{c0}K^0_S}$      &     &      &      &      &      &      &      &      &      & 1.00 \\
 \hline
\end{tabular}
\label{table:full_cor_matrix1_1}
\end{center}
\end{table}

%\begin{table}[htbp]
\begin{table}
\begin{center}
\caption{Statistical correlation matrix for Solution1.}
\begin{tabular}{l|ccccc cccc} 
\hline \hline
& $b_{(K^0_SK^+)_{\rm NR}}$ & $b_{(K^0_SK^-)_{\rm NR}}$ & $d_{f_0(980) K^0_S}$ & $d_{\phi(1020) K^0_S}$ & $d_{\rm others}$ & $c_{f_0(980) K^0_S}$ & $c_{\phi(1020) K^0_S}$ & $c_{\rm others}$ & $\alpha$ \\ \hline
$a_{f_0(980)}$             & 0.02 &-0.09 & 0.11 & 0.01 & 0.14 &-0.20 &-0.05 &-0.01 &-0.35 \\
$a_{\phi(1020) K^0_S}$     & 0.15 &-0.24 & 0.22 & 0.07 & 0.07 & 0.02 & 0.01 & 0.07 &-0.72 \\
$a_{f_{X} K^0_S}$          & 0.11 &-0.28 & 0.15 & 0.08 & 0.06 & 0.01 & 0.01 & 0.05 &-0.65 \\
$a_{\chi_{c0}K^0_S}$       & 0.17 &-0.12 & 0.19 & 0.06 & 0.08 & 0.00 & 0.02 & 0.05 &-0.54 \\
$a_{(K^0_SK^+)_{\rm NR}}$  & 0.26 &-0.25 & 0.20 & 0.08 & 0.26 &-0.14 &-0.01 &-0.05 &-0.81 \\
$a_{(K^0_SK^-)_{\rm NR}}$  &-0.16 &-0.50 & 0.22 & 0.07 & 0.19 &-0.09 &-0.01 &-0.05 &-0.77 \\
$b_{f_0(980)}$             &-0.21 & 0.29 &-0.30 &-0.11 &-0.03 &-0.09 &-0.03 &-0.16 & 0.72 \\
$b_{\phi(1020) K^0_S}$     &-0.20 &-0.01 &-0.00 &-0.09 & 0.11 &-0.20 &-0.03 &-0.09 & 0.15 \\
$b_{f_{X} K^0_S}$          &-0.09 & 0.01 &-0.10 &-0.04 &-0.02 &-0.00 &-0.01 &-0.03 & 0.21 \\
$b_{\chi_{c0}K^0_S}$       & 0.00 & 0.18 & 0.11 & 0.05 & 0.17 &-0.07 &-0.01 & 0.01 &-0.07 \\
$b_{(K^0_SK^+)_{\rm NR}}$  & 1.00 & 0.03 & 0.01 &-0.01 &-0.05 & 0.02 & 0.02 & 0.05 &-0.30 \\
$b_{(K^0_SK^-)_{\rm NR}}$  &      & 1.00 &-0.08 &-0.02 &-0.00 & 0.01 & 0.02 & 0.02 & 0.63 \\ 
$d_{f_0(980) K^0_S}$       &      &      & 1.00 & 0.60 & 0.59 & 0.14 &-0.02 & 0.47 &-0.20 \\
$d_{\phi(1020) K^0_S}$     &      &      &      & 1.00 & 0.35 & 0.45 & 0.08 & 0.37 &-0.07 \\
$d_{\rm others}$           &      &      &      &      & 1.00 &-0.39 &-0.09 &-0.13 &-0.14 \\
$c_{f_0(980) K^0_S}$       &      &      &      &      &      & 1.00 & 0.24 & 0.60 & 0.07 \\
$c_{\phi(1020) K^0_S}$     &      &      &      &      &      &      & 1.00 & 0.14 & 0.00 \\ 
$c_{\rm others}$           &      &      &      &      &      &      &      & 1.00 & 0.00 \\ 
$\alpha$                   &      &      &      &      &      &      &      &      & 1.00 \\ \hline
\end{tabular}
\label{table:full_cor_matrix1_2}
\end{center}
\end{table}

%\begin{figure}[htbp]

%${\bf systematic uncertainty}$
The sources of systematic uncertainties and their contributions are
summarized in Table~\ref{table:systematic-errors1}.
The systematic errors in the vertex reconstruction include uncertainties
in the IP constraint, charged track selection based on track helix
errors, vertex reconstruction quality, $\Delta t$ requirement, tracking
error corrections, $\Delta z$ bias, and imperfect SVD alignment.
The parameters for flavor tagging and resolution function, physics
parameters, background $\Delta t$ shape and signal probability are
varied by $\pm 1 \sigma$.
For each histogram, systematic errors are estimated
using 100 sets of pseudo-experiments, generated by statistically
fluctuating samples to create the histogram.
Samples of pseudo-experiments showed some fitting bias for $CP$ parameters
due to low statistics in each sample.
We take this bias as a systematic uncertainty.
The effect of misreconstruction is accounted for by comparing the fitted
results of signal MC samples with and without misreconstructed events.
The efficiency histogram also includes systematic uncertainties in
the Dalitz-dependent correction factors for $K^0_S$, PID and tracking efficiency.
For tag-side interference~\cite{ref:TSI}, pseudo-experiments are generated with
and without tag-side interference and the difference is taken as a
systematic error.
The fixed masses and widths of the resonance form factors in the signal model are varied
by their errors, the changes in the results are taken into account in the systematic errors.
To take into account a systematic uncertainty due to mass resolution
in the $\phi(1020)$ mass region, the width of the $\phi(1020)$ is varied
from 4.26 MeV/$c^2$ to 5.4 MeV/$c^2$ and the difference in the fitted result is taken.
The systematic uncertainty due to
Blatt-Weisskopf barrier factors~\cite{ref:BWBF}
in RBW is determined by taking the difference in
the fitted results with and without these factors.
The shape of the non-resonant component is empirically chosen, so
different parameterizations are possible.
This includes modeling the non-resonant part with the tail of a Breit-Wigner, 
$R_{NR}(s;\alpha)=i{\alpha}/{(s+i\alpha)}$
and a power law whose exponent is a fitted parameter, $R_{NR}(s;\alpha)=s^{-\alpha}$.
A possible variation in the model of the $f_0(980)$ Flatt\'{e} function, is
also considered using a different parameterization~\cite{ref:flattethree}.
We also include a possible contribution from the spin 2 $f_2(1270)$ resonance in the signal model.
The differences in the fit results from these alternate Dalitz plot parameterizations were summed in quadrature.
The total systematic uncertainty is obtained by summing all of the above
contributions in quadrature.
%\begin{table}[hbtp]
\begin{table}
\begin{center}
\caption{Summary of systematic uncertainties for Solution 1.}
\begin{tabular}{l|ccc|ccc}
\hline\hline & $f_0(980) K^0_S$ & ${\phi(1020) K^0_S}$ &others & $f_0(980) K^0_S$ & ${\phi(1020)
K^0_S}$ &others \\ \hline Category & & $\delta \phi_1^{\rm eff}(^{\circ})$ & & & $\delta
{\cal A}_{CP}$ & \\ \hline \hline 
Vertex Reconstruction          & 1.3 & 1.2 & 1.1 & 0.046 & 0.080 & 0.024   \\
Wrong tag fraction             & 0.2 & 0.2 & 0.2 & 0.004 & 0.006 & 0.003   \\ 
$\Delta t$ resolution function & 1.9 & 1.9 & 1.5 & 0.018 & 0.011 & 0.010   \\
Possible fit bias              & 2.2 & 0.9 & 0.4 & 0.067 & 0.008 & 0.026   \\ 
Physics parameters             & 0.1 & 0.0 & 0.1 & 0.002 & 0.001 & 0.001   \\
Background PDF                 & 1.0 & 0.8 & 0.8 & 0.037 & 0.012 & 0.016   \\ 
Signal fraction                & 0.2 & 0.4 & 0.3 & 0.013 & 0.006 & 0.004   \\ 
Misreconstruction              & 0.1 & 0.0 & 0.0 & 0.000 & 0.000 & 0.001 \\ 
Efficiency                     & 0.2 & 0.2 & 0.1 & 0.011 & 0.004 & 0.005   \\
Signal model                   & 0.7 & 0.4 & 0.4 & 0.040 & 0.017 & 0.006   \\
Tag-side interference          & 0.0 & 0.0 & 0.0 & 0.043 & 0.054 & 0.066 \\\hline 
Total w/o Dalitz model         & 3.4 & 2.6 & 2.1 & 0.110 & 0.100 & 0.078 \\ \hline 
Dalitz model                   & 4.0 & 1.4 & 2.5 & 0.089 & 0.019 & 0.032\\ \hline \hline
%Total (0-11)                       &&&&&&\\ \hline \hline 
\end{tabular}
\label{table:systematic-errors1}
\end{center}
\end{table}

%KM Summary
%${\bf Summary}$
In summary, for the first time in Belle we perform a measurement
of the $CP$ violating asymmetries in $B^0(\overline{B}^0) \rightarrow K^+K^-K^0_S$ decays
with the time-dependent Dalitz approach. 
There are four solutions that describe the data well.
These give similar values for the $CP$ violating phase in the $b \rightarrow s$ penguin mode,
$B^0\to \phi(1020) K^0_S$,
\begin{eqnarray*}
\phi_1^{\rm eff}(B^0\to \phi(1020) K^0_S) & = & (32.2 \pm 9.0 \pm 2.6 \pm 1.4)^{\circ};\\
\phi_1^{\rm eff}(B^0\to \phi(1020) K^0_S) & = & (26.2 \pm 8.8 \pm 2.7 \pm 1.2)^{\circ};\\
\phi_1^{\rm eff}(B^0\to \phi(1020) K^0_S) & = & (27.3 \pm 8.6 \pm 2.8 \pm 1.3)^{\circ}\; {\rm and}\\
\phi_1^{\rm eff}(B^0\to \phi(1020) K^0_S) & = & (24.3 \pm 8.0 \pm 2.9 \pm 5.2)^{\circ}.
\end{eqnarray*}
These $CP$ violating parameters are consistent at the current level of precision
with the measurement
of the $CP$ violating phase in $b\to c \bar{c}s$ processes such as $B^0\to J/\psi K^0$,
which is $(22\pm 1)^{\circ}$.
Previous measurements used $K^+K^-$ selections around the $\phi$(1020) mass and
in the higher $K^+K^-$ mass region.
Here we establish a superior analysis procedure for obtaining $CP$ violating 
asymmetries without the uncertainty from interference among different resonant contributions, and therefore
this represents an important step toward measurements with
higher statistics such as in Super B-factory experiments.
%~\cite{belleII}.

%----------- Long version, for most papers ----------- 
We thank the KEKB group for the excellent operation of the
accelerator, the KEK cryogenics group for the efficient
operation of the solenoid, and the KEK computer group and
the National Institute of Informatics for valuable computing
and SINET3 network support.  We acknowledge support from
the Ministry of Education, Culture, Sports, Science, and
Technology (MEXT) of Japan, the Japan Society for the 
Promotion of Science (JSPS), and the Tau-Lepton Physics 
Research Center of Nagoya University; 
the Australian Research Council and the Australian 
Department of Industry, Innovation, Science and Research;
the National Natural Science Foundation of China under
contract No.~10575109, 10775142, 10875115 and 10825524; 
the Ministry of Education, Youth and Sports of the Czech 
Republic under contract No.~LA10033 and MSM0021620859;
the Department of Science and Technology of India; 
the BK21 and WCU program of the Ministry Education Science and
Technology, National Research Foundation of Korea,
and NSDC of the Korea Institute of Science and Technology Information;
the Polish Ministry of Science and Higher Education;
the Ministry of Education and Science of the Russian
Federation and the Russian Federal Agency for Atomic Energy;
the Slovenian Research Agency;  the Swiss
National Science Foundation; the National Science Council
and the Ministry of Education of Taiwan; and the U.S.\
Department of Energy.
This work is supported by a Grant-in-Aid from MEXT for 
Science Research in a Priority Area (``New Development of 
Flavor Physics''), and from JSPS for Creative Scientific 
Research (``Evolution of Tau-lepton Physics'').


\begin{thebibliography}{99}
\bibitem{phi1_beta}
Another naming convention $\beta(=\phi_1)$ is also used in literatures.
\bibitem{charge_conj} Throughout this paper, the inclusion of the charge-conjugate decay mode
is implied unless otherwise stated.
\end{thebibliography}
\end{document}